\documentclass[twocolumn,twocolappendix]{aastex631}
\usepackage{subfigure}
\usepackage[colorinlistoftodos]{todonotes}
\usepackage{amssymb}
\usepackage{comment}
\usepackage{multirow,acronym}
\usepackage{natbib}
\usepackage{makecell}
\usepackage{booktabs}
\usepackage{amsmath}
\usepackage{fancyvrb}
\usepackage{color}
\usepackage{xcolor}
\usepackage{hyperref}
\usepackage[T1]{fontenc}
\usepackage{graphicx}
\usepackage{bm}
\usepackage{float}
\usepackage{aas_macros}
\usepackage{float}
\usepackage{amsmath,amssymb}
\usepackage{natbib} 
\usepackage{graphicx} 
\usepackage{color}
\usepackage{verbatim}
\usepackage{rotating}

\newcommand{\proptosim}{\mathrel{\vcenter{
 \offinterlineskip\halign{\hfil$##$\cr
 \propto\cr\noalign{\kern2pt}\sim\cr\noalign{\kern-2pt}}}}}
\renewcommand{\b}[1]{\boldsymbol{#1}}
\renewcommand{\i}{\ensuremath{\rm i}}
\newcommand{\unit}[1]{{\rm\, #1}}

\newcommand{\m}{\unit{m}}
\newcommand{\g}{\unit{g}}
\newcommand{\G}{\unit{G}}
\newcommand{\K}{\unit{K}}

\newcommand{\s}{\mathrm{s}}

\newcommand{\B}{\mathbf{B}}     
\newcommand{\E}{\mathbf{E}}     
\newcommand{\J}{\mathbf{J}}     
\newcommand{\F}{\mathbf{F}}     
\renewcommand{\v}{\mathbf{v}}   

\renewcommand{\d}{\mathrm{d}}
\newcommand{\e}{\mathrm{e}}

\newcommand{\p}{\mathrm{p}}     

\renewcommand{\O}{\mathrm{O}}     
\newcommand{\A}{\mathrm{A}}     
\renewcommand{\H}{\mathrm{H}}     
\renewcommand{\P}{\mathrm{P}}     

\renewcommand{\i}{\ensuremath{\mathrm{i}}}
\newcommand{\twinkle}{\texttt{Twinkle} }
\newcommand{\vbbl}{\texttt{VBBL} }

\def\E{{\rm E}}
\def\s{{\rm S}}

\renewcommand{\l}{{\rm L}}

\begin{document}

\title {\texttt{Twinkle}: A GPU-based binary-lens microlensing code with
  contour integration method}

\author{Suwei Wang}
\affil{The Kavli Institute for Astronomy and Astrophysics,
  Peking University, Beijing 100871, China}
\affil{Department of Astronomy, School of Physics, Peking
  University, Beijing 100871, China}
  
\author{Lile Wang}
\affil{The Kavli Institute for Astronomy and Astrophysics,
  Peking University, Beijing 100871, China}
\affil{Department of Astronomy, School of Physics, Peking
  University, Beijing 100871, China}

\author{Subo Dong}
\affil{Department of Astronomy, School of Physics, Peking
  University, Beijing 100871, China}
\affil{The Kavli Institute for Astronomy and Astrophysics,
  Peking University, Beijing 100871, China}
\affil{National Astronomical Observatories, Chinese Academy of Science, Beijing 100101, China\\}

\correspondingauthor{Suwei Wang} 
\email{suwei\_wang@stu.pku.edu.cn}

\correspondingauthor{Lile Wang} 
\email{lilew@pku.edu.cn}

\correspondingauthor{Subo Dong} 
\email{dongsubo@pku.edu.cn}

\begin{abstract}
With the rapidly increasing rate of microlensing planet detections, microlensing modeling software faces significant challenges in computation efficiency. Here, we develop the \twinkle code, an efficient and robust binary-lens modeling software suite optimized for heterogeneous computing devices, especially GPUs. Existing microlensing codes have the issue of catastrophic cancellation that undermines the numerical stability and precision, and \twinkle resolves them by refining the coefficients of the binary-lens equation. We also devise an improved method for robustly identifying ghost images, thereby enhancing computational reliability. We have advanced the state of the art by optimizing \twinkle specifically for heterogeneous computing devices by taking into account the unique task and cache memory dispatching patterns of GPUs, while the compatibility with the traditional computing architectures of CPUs is still maintained. \twinkle has demonstrated an acceleration of approximately $2$ orders of magnitude ($\gtrsim 10^2\times$) on contemporary GPUs. The enhancement in computational speed of  \twinkle will translate to the delivery of accurate and highly efficient data analysis for ongoing and upcoming microlensing projects. Both GPU and CPU versions of \twinkle are open-source and publicly available.
\end{abstract}


\section{Introduction}

Gravitational microlensing is one of the main methods for detecting extrasolar planets (hereafter exoplanets), 
and the method is most sensitive to exoplanets near or beyond the snow line \citep{1991ApJ...374L..37M, 1992ApJ...396..104G, 2012ARA&A..50..411G}. In recent years, thanks to projects such as the Optical Gravitational Lensing Experiment IV \citep{Udalski2015}, the Korea Microlensing Telescope Network (\citealt{Kim2016}), and the Microlensing Observations in Astrophysics (\citealt{MOA}) , the microlensing planet discovery rate has been greatly boosted \citep[see recent reviews by][]{2021ARA&A..59..291Z, 2023arXiv231007502M}. The upcoming space-based projects, such as the Nancy Grace Roman Space Telescope (previously known as WFIRST), are expected to significantly expand the yield and discovery space \citep{Penny2019, Yee2023}.

With the increasing yields of surveys, planetary microlensing analysis faces pressing computational challenges. Light-curve modeling of a star–planet lens system requires solving the two-point-mass-lens (i.e., binary-lens) equation \citep{SchneiderWeiss1986, Witt1990}  to calculate the source star's magnification. Planetary microlensing modeling often needs to compute the finite-source effects \citep{Gould1994, NemiroffWickramasinghe1994, WittMao1994}, which are far more time-consuming than point-source computations. Enhancing the speed of finite-source binary-lens computation is at the core of improving the efficiency of planetary microlensing analysis. 

Due to the conservation of the surface brightness in lensing \citep{Misner1973}, the magnification is equal to the ratio between the total area of all images and the area of the source for a uniform source. There are two main categories of algorithms, namely, the ray-shooting and the contour integration methods, which differ in how the image area is calculated.

In the ray-shooting method \citep{1986A&A...166...36K, SchneiderWeiss1986, 1997MNRAS.284..172W}, 
the image plane is sampled on grids from which rays are traced back to the source plane. The magnification is determined by counting the number of rays falling in the source. The one-to-one image-to-source mapping is easy to compute, and it is straightforward to account for the limb-darkening effects by simply weighting each ray with the brightness at the corresponding source position. A ray-shooting map can be saved for repetitive usage to ease the parameter space exploration of a static binary-lens model. However, highly magnified images are computationally demanding for ray shooting. A main limitation of the standard ray-shooting method is the lack of flexibility when modeling the orbital motion of the lens system. Improved ray-shooting algorithms have been developed to address the drawbacks of the standard approach  \citep{2006ApJ...642..842D, 2009ApJ...698.1826D, 2010ApJ...716.1408B}.

In the contour integration method, discrete points along the source's boundary are mapped onto the image plane by solving the binary-lens equation. Then, the image contours are constructed by connecting adjacent points on the image plane \citep{1987A&A...174..361S, 1995A&AS..109..597D}. Integrations along these contours yield the magnifications through the Stokes theorem \citep{1997ApJ...477..580G, 1998A&A...333L..79D}. Compared with ray shooting, the contour integration method is much more flexible, making it particularly suitable for modeling lens orbital motion. However, it is generally much more time-consuming to compute a limb-darkened source than a uniform one. Connecting points to construct image contours can sometimes be complicated (e.g., around a cusp) and needs careful treatments. Variants of this method have been developed to overcome these drawbacks \citep[e.g.,][]{2006ApJ...642..842D, 2007MNRAS.377.1679D}. Contemporarily, the most widely used contour integration
algorithm is VBBinaryLensing (hereafter \vbbl; \citealt{2010MNRAS.408.2188B}, \citealt{2018MNRAS.479.5157B}), which has significant improvement in speed and reliability by introducing parabolic corrections to the integrals and an analytical estimator for the error control and optimizations over the contour sampling. 

In this work, we present \texttt{Twinkle}\footnote{\href{https://github.com/AsterLight0626/Twinkle}{https://github.com/AsterLight0626/Twinkle}}, a contour integration code with improved performance in efficiency and robustness. \twinkle incorporates optimizations on multiple fronts, ranging from the numerical evaluation of lens equation coefficients to the solution of the algebraic equations with high concurrency.  Our primary goal in developing \twinkle is to unleash the computational capabilities of high-performance heterogeneous devices, especially GPUs. 

This paper is structured as follows. \S~\ref{sec:method} discusses the methods used in \texttt{Twinkle}, including optimization in polynomial coefficient evaluation, a numerically robust criterion
for determining the existence of ghost images, and an adaptive sampler of contour integrations for GPUs. \S~\ref{sec:speed} compares the performance of the \twinkle code on GPUs with \vbbl on CPUs.

\section{Methods}
\label{sec:method}

\subsection{Lens equation}
\label{subsec:basic-equations}

Solving the binary-lens equation for point sources to form the image contour is at the basis of the contour integration method. The calculations on different points are independent, for which parallelization can be easily implemented for GPUs.

We adopt the coordinate system convention that the origin is fixed on the center of mass of the binary-lens system. This convention leads to the following form of lens equations using complex variables,
\begin{equation}
  \label{eq:method-lens-eq}
    \zeta = z + \dfrac{m_1}{\bar{z}_1-\bar{z}} +
    \frac{m_2}{\bar{z}_2-\bar{z}},
\end{equation}
where the subscripts denote the binary objects (1 for the primary or the host star, 2 for the secondary or the planet); $m_{1,2}$ stand for the mass of binary components normalized by their total mass ($M_{\rm tot}$); $z$ and $\zeta$ are complex variables describing the
positions of the images and sources, respectively; the overbar (e.g. $\bar{z}$) denotes complex
conjugate; and $z_{1,2}$ are the binary positions. 
Using two binary-lens parameters $s$, the binary's projected separation normalized by the Einstein radius, and $q\equiv m_2/m_1$, the binary mass ratio, the complex variables $z_{1,2}$ can be expressed as
\begin{equation}
  z_1 = - \frac{q s}{1+q}, \quad  z_2 = \frac{s}{1+q}.
\end{equation}
Note that all complex coordinates ($\zeta$, $z$ and $z_{1,2}$) represent angular positions normalized by the angular Einstein radius,
\begin{equation}
\label{eq:method-eistein-radius}
\theta_{\rm E} = \left(\dfrac{4GM_{\rm tot}}{c^2}
\dfrac{D_\s - D_\l}{D_\l D_\s}\right)^{1/2},
\end{equation}
in which $D_{\rm \l}$ and $D_{\rm \s}$ are the distances between the observer and the lens (subscript ``L'') and source (``S''), respectively.
The computational cost of the contour integration method is dominated by repeatedly solving Equation~\eqref{eq:method-lens-eq}, and we discuss its optimizations in the following sections.

\subsection{Coefficients of the lens equation}\label{sec:coeff_generator}
Equation~\eqref{eq:method-lens-eq} is solved numerically after transformation into a fifth-order polynomial equation. We adopt the algorithm developed by  \citet{2012arXiv1203.1034S}, a combination of the Newton–Raphson method with second-order
corrections and Laguerre iterations, and it is implemented in the solver of \vbbl \citep{2010MNRAS.408.2188B,
  2018MNRAS.479.5157B}. 
Although solving the binary-lens equation seems well established, we identify numerical issues in the existing implementations.
These issues could introduce seriously inaccurate numerical coefficients when $q \ll 1$, undermining  the analysis for planets with Earth mass or below  ($q\lesssim 10^{-5}$) at the frontier of microlensing exoplanet searches. Now consider the conventional  polynomial form of the
lens equation used by the \vbbl program, for which the origin is moved to $z_2$ and new coordinates $y, \eta$ are defined as follows:
\begin{equation}
  \label{eq:lens-poly-orig}
  \begin{split}
    y \equiv & \zeta - z_2, \
    \eta \equiv z - z_2, \
    0 = \sum_{i=0}^{5}c_i \eta^i; \\
    c_5 = & -\bar{y}(\bar{y} + s),\\ 
    c_4 = &\bar{y} \left[\bar{y}(-2s+y) -1 \right]  +s \left[\bar{y}(-2s+y) -m_1\right] , \\
    c_3 = & 2\bar{y}y - s^3\bar{y} + s^2 \left[\bar{y} (2y-\bar{y}) - m_1 \right] \\ 
    &-a\left[y + 2\bar{y}(\bar{y}y - m_2) \right], \\
    c_2 = & y(1+s^3\bar{y}) + s\left[m_1+2y\bar{y}(1+m_2)\right] \\
    &+s^2\left[\bar{y}(m_1-m_2) + y(1+m_2+\bar{y}^2)\right], \\
    c_1 = & m_2 s \{ s \left[ m_1 + y(2\bar{y} + s) \right ] + 2y \}, \\
    c_0 = & s^2 m_2^2 y. \\
  \end{split}
\end{equation}

The numerical subtractions of terms in the $\{c_i\}$ polynomial
coefficient expressions often lead to results proportional
to $q^n$ ($n\in \mathbb{N}^+$); thus, if implemented directly
in a code, it could lead to the classic ``catastrophic cancellation'' 
issue in floating point arithmetic of losing accuracy when subtracting 
numbers of similar magnitudes. This issue could cause significantly large 
error in coefficient evaluation, undermining the accuracy and 
numerical stability of solutions. This catastrophic cancellation
issue has been overlooked in existing microlensing modeling codes.

We solve the issue of catastrophic cancellation by refactorizing the
coefficients. For instance, the following simple equality,
\begin{equation}
  \label{eq:method-example-factorize}
  1 + 2q - (1+q)^{-1} \equiv q \left[2 + (1+q)^{-1} \right],
\end{equation}
recasts  $[1 + 2q - (1+q)^{-1}]$, a term commonly appearing in 
Equation~\eqref{eq:lens-poly-orig}, into a form that fully avoids the 
subtraction of numbers that are relatively large compared to $q$. 
To facilitate refactorizing, we introduce two intermediate variables 
$(v_{c}, v_{p})$ and recast Equation~\eqref{eq:lens-poly-orig} as
\begin{equation}
  \label{eq:method-transformed}
  \begin{split}
  y \equiv & \zeta - z_2, \
    \eta \equiv z - z_2, \
    0 = \sum_{i=0}^{5}c_i \eta^i; \\
  v_c \equiv &\bar{y} + s , \
  v_p \equiv 1+s\bar{y}; \\
  c_5 = &-\bar{y}v_c, \
  c_4 = (\bar{y}y - 1 - 2s\bar{y}) v_c +m_2s,  \\
  c_3 = &(2y-s)\left[ v_p v_c
          + 2{\mathrm{Im}(}y{)\mathrm{i}} - m_2s     \right]  \\
          &+ 4(m_2s - y){\mathrm{Im}(}y{)\mathrm{i}}  , \\
  c_2 = & \mathrm{Re}(v_p)^2
    \mathrm{Re}(v_c)    \\
    &+\mathrm{Im}( y ) \mathrm{Re}(v_p)
    \mathrm{Re}( 1-sy )\mathrm{i}
    \\
    & + s{\mathrm{Im}(} y {)^2} ( 2 + s v_c)    \\
    & + m_2s \left[
    2\bar{y}(y-s) - (1-sy)
    \right],
    \\
  c_1 = & m_2s\left[ (s+2y)(v_p)
    + s(2s{\mathrm{Im}(}y{)\mathrm{i}} - m_2)  \right],    \\
  c_0 = & s^2m_2^2y, \\
    \end{split}
\end{equation}
  
where Re and Im denote the real and imaginary parts of complex numbers. Compared to the original Equation~\eqref{eq:lens-poly-orig}, our modified Equation~\eqref{eq:method-transformed} avoids subtractions with high-order terms of $q$. For example, when the source crosses a planetary caustic, $v_p \approx \mathrm{Im}(y) \approx \sqrt{q}$ \citep{2006ApJ...638.1080H}, resulting in $c_2 \approx q$. In Equation~\eqref{eq:method-transformed}, all terms in $c_2$ are on the order of $q$, while this is not the case for Equation~\eqref{eq:lens-poly-orig}.

\begin{figure*}
\centering
\vspace{-1.5cm}
\includegraphics[width=17cm]{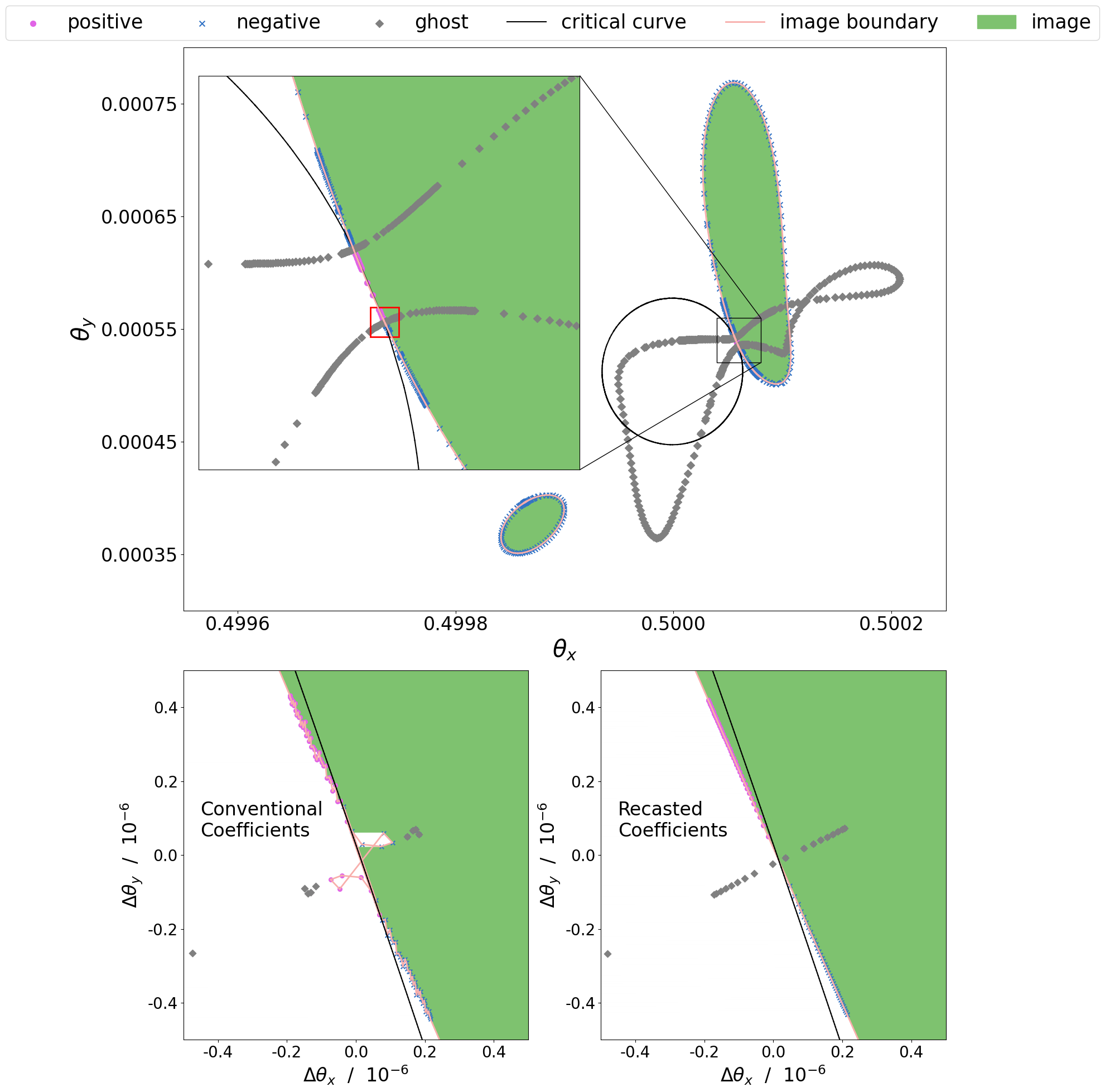}

\caption{An example demonstrating the enhanced accuracy introduced by the improvements in generating the coefficients of the lens equation 
  (Equation~\ref{eq:lens-poly-orig} in \S\ref{subsec:basic-equations}). The top panel shows the overall images (green) on the image plane ($\theta_x$-$\theta_y$). The bottom left and right panels are the
  zoom-ins of the region marked by the square in the top
  panel, showing the results using conventional (bottom left panel;
  Equation~\ref{eq:lens-poly-orig}) and our improved (bottom right
  panel; Equation~\ref{eq:method-transformed}) schemes of
  generating the coefficients. In the bottom panels, the origin is shifted to the central position of the zoom-in region, with  $\Delta \theta_x = (\theta_x - 0.5000595)$ and $\Delta \theta_y = (\theta_y - 0.000535)$, where the real images and ghost images (gray diamonds) mapped from the source boundary cross on the critical curve (black solid line). The real images on the boundary with positive and negative parities are marked with magenta dots and blue crosses, respectively. The pink lines connecting real images form the boundary of the extended image for the contour integration method. As shown in the bottom panels, the improved accuracy of \twinkle's new method yields a more accurate image contour and hence more reliable magnification than conventional method.}
\label{fig:coeff_comp}
\end{figure*}

We demonstrate the improvement in numerical accuracy of Equation~\eqref{eq:method-transformed} over Equation~\eqref{eq:lens-poly-orig} using an example presented in Figure~\ref{fig:coeff_comp}. For this example, we use a set of binary-lens parameters $(s = 0.5, q = 10^{-6})$ for a low-mass planetary system. We compute points on a circular source with radius $\rho = 10^{-4}$ normalized by the Einstein radius, and the source centers at $\zeta =  -1.49981640625+0.0035 \mathrm{i}$, where the source boundary is close to a cusp of the planetary caustic. The bottom right panel uses coefficients from Equation \ref{eq:method-transformed}, while the bottom left panel uses the method from Equation \ref{eq:lens-poly-orig}. The inaccurate image positions derived from the conventional coefficient-generation method result in incorrect image morphology and hence erroneous magnification, whereas the recasted coefficient results are accurate. 

\subsection{Image Construction}\label{sec:image construction}

Before performing finite-source calculations, \twinkle checks the adequacy of the point-source approximation by adopting the methods outlined by \cite{2018MNRAS.479.5157B}. These include the quadrupole test \citep{Gould2008, Pejcha2009, Cassan2017} and the ghost images test. Computational time can be significantly reduced if the point-source approximation is adequate.

\twinkle uses the contour integration method to calculate the finite-source effect. The boundary of the source is circular, with a normalized radius $\rho$. Points on the boundary are parameterized by the angle $\theta$, measured counterclockwise from the positive $x$-axis. This angle serves as an ordering parameter for sorting sampling points along the boundary.

Each point source produces five images from Equation~\eqref{eq:method-transformed}, divided into three classes: positive, negative, and ghost. 
Ghost images do not satisfy lens equation \eqref{eq:method-lens-eq}, while real images satisfy and are divided by the parity into positive and negative. Parity $p$ is the symbol of Jacobian $J(z)$, 
\begin{equation}\label{eq:jacobian}
\begin{aligned}
    &f(z) \equiv \frac{m_1}{{z}_1-{z}} + \frac{m_2}{{z}_2-{z}}, f'(z) \equiv \frac{\mathrm{d}f(z)}{\mathrm{d}z}, \\
    & J(z) = 1-|f'(z)|^2 , \\
    & p=
\begin{cases}
+1, & J > 0\\
-1, & J < 0
\end{cases}
,
\end{aligned}
\end{equation}
defined for each real image.

All real images are connected to the images with the same parity and from adjacent point sources unless cross-caustic occurs here. For a lens system with $N$ lens objects, there are always $N-1$ more negative images than positive images \citep{2001astro.ph..3463R}. For a binary system, parity sum is always $-1$, which means one positive with two negative images or two positive with three negative images. If multiple images are connected, e.g., three images connect to three images, the way to minimize the sum of all distances is selected.

The numbers of real images differ in adjacent point sources when cross-caustic occurs. In that case, the additional real image pair from the same source connects. For example, $\theta_i$ has three real images, and $\theta_{i+1}$ has five real images. After three real images in $\theta_i$ find their partner, one positive and one negative image in $\theta_{i+1}$ are left, which connect to each other.

\begin{figure*}
\centering
\hspace{-0.8cm}
\includegraphics[width=16cm]{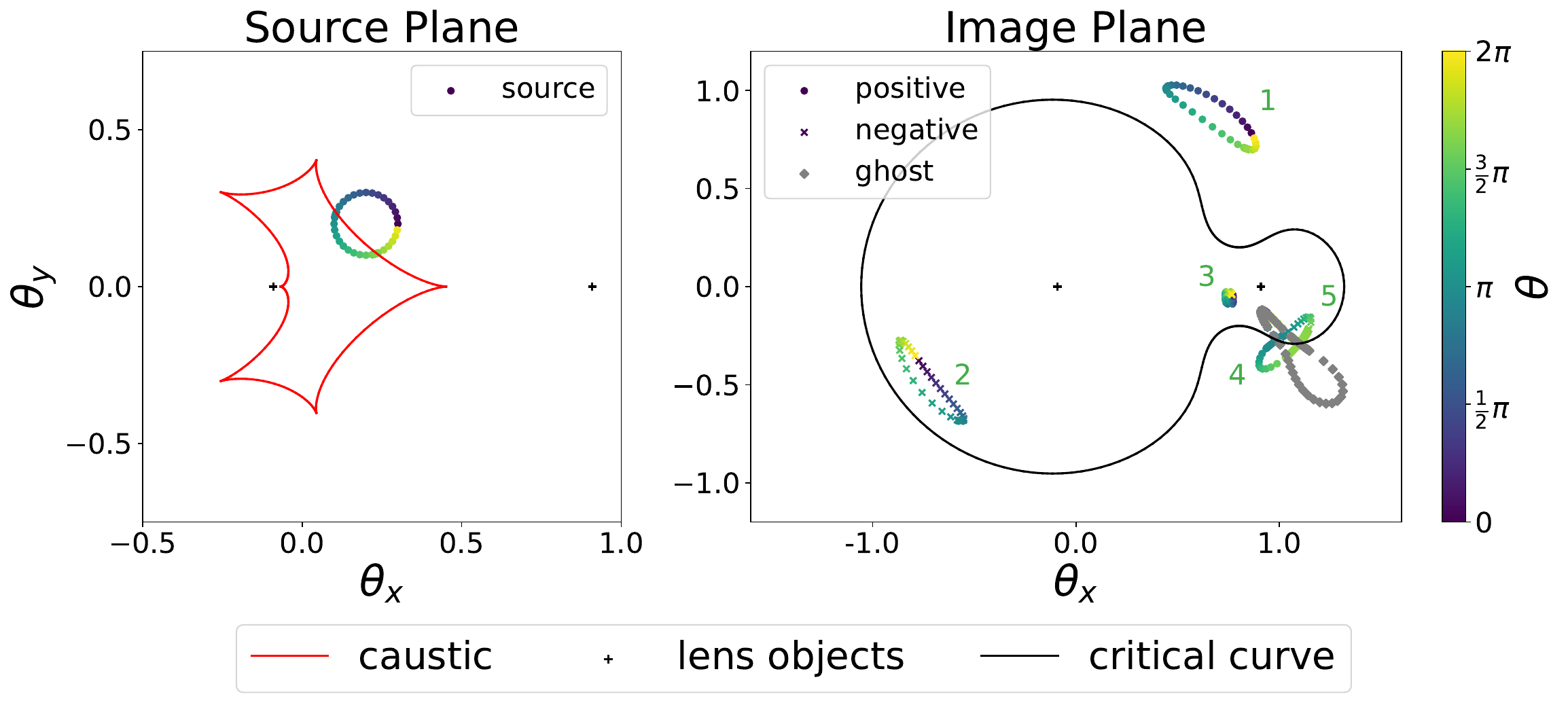}
\caption{Visualization of an extended source and the images. The left and right panels show the source and image planes, respectively. The caustic is shown in red, and the critical curve is displayed in black. The two black plus signs mark the positions of two lens objects. The source boundary in the left panel crosses the caustic twice, creating three complete image boundaries ($1$, $2$, and $3$ in the right panel) and two broken tracks ($4$ and $5$). The points with the \textit{viridis} color map in the right panel are real images belonging to the point sources with the same colors. The colors represent the source position angle $\theta$. The real images inside the critical curve have negative parities and are marked by crosses, while the images outside (dots) have positive parities. The gray diamonds represent ghost images generated by points on the source boundary located outside the caustics.}
\label{fig:s1q0.1}
\end{figure*}

Figure \ref{fig:s1q0.1} visualizes the images of an extended source around a caustic. The lens system has mass ratio $q = 0.1$ and binary separation $s = 1$, with the position of the two lens objects marked by black plus signs in both panels. The left panel shows the source plane, in which a source with radius $\rho = 0.1$ centered at $\zeta = 0.2+0.2\mathrm{i}$ crosses the caustics (red). The boundary of the source is plotted using the \textit{viridis} color map, transitioning from dark to light along the counterclockwise position angle, starting from the direction of the $x$ -axis. The right panel shows the corresponding images using the same color map. The real images outside the critical curve (black) have positive parities and are marked as dots,  while those inside the critical curve, having negative parities, are represented by crosses.
Point sources inside caustics produce five real images, while those outside produce three images \citep{1986A&A...164..237S}. Since only part of the source is located inside the caustic, only three complete images are produced, shown in the right panel and marked by $1$, $2$, and $3$. Two ``broken'' tracks, $4$ and $5$, are produced by the part of the source inside the caustic, and they connect and form an image. Based on their parities, tracks with positive parities ($1$ and $4$) follow a counterclockwise direction, the same as the source, while tracks $2$, $3$, and $5$ are clockwise.

After the images connect, the total extended image area $A$ is calculated by Stokes's theorem. The total image area could be divided into small pieces area between each connected image pair $z_{i}$ and $z_{i+1}$,
\begin{equation}
\begin{aligned}
    A &= \sum_I \oint_R \frac{1}{2} \boldsymbol{z} \wedge \mathrm{d} \boldsymbol{z} \\
    &= \sum_i \int_{\boldsymbol{z}_i}^{\boldsymbol{z}_{i+1}} \frac{1}{2} \boldsymbol{z} \wedge \mathrm{d} \boldsymbol{z} ,
\end{aligned}
\end{equation}
where $I$ means the sum for all extended images and $R$ means that the orientation of integration is a right-hand loop. The complex number $\boldsymbol{z} = a+b\mathrm{i}$ is the coordinate of the images, and the wedge product $\boldsymbol{z}_1 \wedge \boldsymbol{z}_2$ is defined as $a_1b_2 - a_2b_1$. The integration in the second row is along the boundary of the extended image, where $\boldsymbol{z}_i$ and $\boldsymbol{z}_{i+1}$ are the coordinates of the connected image pairs.

For each small pieces area between $z_{i}$ and $z_{i+1}$, \twinkle adopts the trapezium approximation $\mathrm{d}A^{(t)}$ and the parabolic correction $\mathrm{d}A^{(p)}$. The trapezium approximation $\mathrm{d}A^{(t)}$ is the area of the triangle formed by connected images and the origin point,
\begin{equation}
\begin{aligned}
    &\int_{\boldsymbol{z}_i}^{\boldsymbol{z}_{i+1}} \frac{1}{2} \boldsymbol{z} \wedge \mathrm{d} \boldsymbol{z} \approx \mathrm{d}A^{(t)} +  \mathrm{d}A^{(p)} \\
    &\mathrm{d}A^{(t)} = \frac{1}{2} \boldsymbol{z}_i \wedge (\boldsymbol{z}_{i+1} - \boldsymbol{z}_i) = \frac{1}{2} \boldsymbol{z}_i \wedge \boldsymbol{z}_{i+1} .
\end{aligned}
\end{equation}
The parabolic correction $\mathrm{d}A^{(p)}$ takes the form from \cite{2010MNRAS.408.2188B} and \cite{VBBL4}.  For usual images, there are two types of parabolic correction $\mathrm{d}A^{(p_1)}$ and $\mathrm{d}A^{(p_2)}$,
\begin{equation} \label{area_cal}
\begin{aligned}
    \mathrm{d}A^{(p_1)} &= \frac{1}{24} \left [ (\boldsymbol{z}_i' \wedge \boldsymbol{z}_i'') + (\boldsymbol{z}_{i+1}' \wedge \boldsymbol{z}_{i+1}'') \right ] \Delta \theta^3  ,  \\
    \mathrm{d}A^{(p_2)} &= \frac{1}{12} (\boldsymbol{z}_{i+1} - \boldsymbol{z}_i) \wedge (\boldsymbol{z}'_{i+1} - \boldsymbol{z}'_i)   \Delta \theta  ,   \\
    \boldsymbol{z}' \wedge \boldsymbol{z}'' &= \left \{ \rho^2 + \mathrm{Im}\left[ (z')^2\zeta' f''(z)  \right]    \right \} ,  \\
    z' &= \left[ \zeta' - f'(z)\bar{\zeta}' \right] J^{-1} , \\
    \zeta' &= \mathrm{i}\rho e^{\mathrm{i}\theta} ,
\end{aligned}
\end{equation}
where $J$ is Jacobian, $f(z)$ is defined in Equation~\eqref{eq:jacobian}, and $\Delta \theta = \theta_{i+1}-\theta_{i}$, in which $\theta_{i+1}$ and $\theta_i$ are source position angles corresponding to $\boldsymbol{z}_{i+1}$ and $\boldsymbol{z}_i$ separately. The final parabolic correction $\mathrm{d}A^{(p)}$ adopts the mean of them as $\mathrm{d}A^{(p)} = (\mathrm{d}A^{(p_1)}+\mathrm{d}A^{(p_2)})/2$. For cross-caustic cases, the parabolic correction is modified, 
\begin{equation}
\begin{aligned}
    \mathrm{d}A^{(p)} &= \frac{1}{24} \left [ (\boldsymbol{z}_+' \wedge \boldsymbol{z}_+'') - (\boldsymbol{z}_-' \wedge \boldsymbol{z}_-'') \right ] \widetilde{\Delta \theta}^3  ,  \\
    \widetilde{\Delta \theta} &= \frac{|\boldsymbol{z}_+ - \boldsymbol{z}_-|}{\sqrt{|\boldsymbol{z}'_+ \cdot \boldsymbol{z}'_-|}} ,
\end{aligned}
\end{equation}
where subscripts $+$ and $-$ represent the positive and negative images. 

\subsection{Ghost Image Detector}\label{sec:ghost_detector}

There are always five solutions produced by polynomial equation \eqref{eq:method-transformed}, but some of them are ghost images that do not satisfy lens equation \eqref{eq:method-lens-eq}. It is necessary to distinguish ghost images from real images that satisfy the lens equation. However, numerical problems make comparing the lens equation residual with $0$ complex. 
The empirical techniques used previously \citep{2018MNRAS.479.5157B, 2021MNRAS.503.6143K} work well in most cases but may fail when the mass ratio $q$ becomes very low.
Therefore, \twinkle develops a new robust ghost image detector consisting of two parts: residual detector and slope detector. The residual detector works for each point source, while the slope detector works only for the point sources near caustics.

\subsubsection{Residual Detector}

There is a particular property of residuals that the sum of all residuals is always equal to $0$. The residual of each root $r_i$ of the polynomial equation could be marked as $R_i$,
\begin{equation}
    R_i = \zeta - r_i + \frac{m_1}{\bar{r}_i - \bar{z}_1} + \frac{m_2}{\bar{r}_i - \bar{z}_2} .
\end{equation}

With the help of Vieta's theorem, the sum of residuals could be simplified. Noting the polynomial equation transformed from the lens equation \eqref{eq:method-lens-eq} as $P(z)$, the residual sum could be written as 
\begin{equation}\label{sum_r}
\begin{aligned}
    P(z) &\equiv \sum_{i=0}^{5}c_i z^i = 0 , \\
    \sum_{i=0}^{5}{R_i} &= 5\zeta - \sum_{i=0}^{5}{r_i} + \sum_{i=0}^{5}{\frac{m_1}{\bar{r}_i - \bar{z}_1}} + \sum_{i=0}^{5}{\frac{m_2}{\bar{r}_i - \bar{z}_2}} \\
    &= 5\zeta + \frac{c_4}{c_5} - m_1\left(\frac{P'(z_1)}{P(z_1)}\right) ^ \dagger - m_2\left(\frac{P'(z_2)}{P(z_2)}\right) ^ \dagger ,
\end{aligned}
\end{equation}
where $P'(z) \equiv d(P(z)) / dz$ is the derivative of the polynomial function, and the dagger symbol $\dagger$ means take the conjugation of the term inside the round brackets.

By substituting the expressions from Equation~\eqref{eq:method-transformed}, it can be proven that the sum of the residuals equals $0$ in this reference frame. Noting that $z_1 = -s$, $z_2=0$, and the source position is denoted as $y$ at this time,
\begin{equation}\label{sum_r2}
\begin{aligned}
    \frac{c_4}{c_5} = &-y + 2s + \frac{\bar{y} + s - m_2 s}{\bar{y}(\bar{y} + s)} ,  \\
    \frac{P'(z_1)}{P(z_1)} =& \frac{-1}{m_1s}(-2s\bar{y} - s^2 + 2) + \frac{m_2}{m_1(s+y)} , \\
    \frac{P'(z_2)}{P(z_2)} = &\frac{1}{m_2s}(2s\bar{y} + s^2 + 2) + \frac{m_1}{m_2y} , \\
    \sum_{i=0}^{5}{R_i} = &5y -y + 2s + \frac{\bar{y} + s - m_2 s}{\bar{y}(\bar{y} + s)} \\
    & \       \ - \left( 4y + 2s + \frac{\bar{y} + m_1s}{\bar{y}(\bar{y}+s)} \right)   \\
    = &0
\end{aligned}
\end{equation}

The result remains valid for any translation and rotation of the reference frame. Considering a translation $\delta$ and rotation $\phi$, the positions are changed to $z_{1,2}' = e^{\mathrm{i}\phi}(z_{1,2}+\delta), \zeta' = e^{\mathrm{i}\phi}(z_{1,2}+\delta), r_{i}' = e^{\mathrm{i}\phi}(r_{i}+\delta)$. The residuals and the sum will change as follows: 
\begin{equation}
\begin{aligned}
    R_{i}' =& e^{\mathrm{i}\phi} (\zeta + \delta - r_i - \delta)   \\
    &+ \frac{m_1}{e^{-\mathrm{i}\phi}(\bar{r}_i+ \bar{\delta} - \bar{z}_1- \bar{\delta})} + \frac{m_2}{e^{-\mathrm{i}\phi}(\bar{r}_i+ \bar{\delta} - \bar{z}_2- \bar{\delta})}     \\
    = & e^{\mathrm{i}\phi} R_{i} , \\
    \sum_{i=0}^{5}{R_i'} &= e^{\mathrm{i}\phi} \sum_{i=0}^{5}{R_i} = 0 .
\end{aligned}
\end{equation}

The residual detector is based on the zero-sum result and the fact that the binary-lens system has two or zero ghost images \citep{1986A&A...164..237S}. If ghost images exist, there must be a pair of ghost images with opposite residuals. On the other hand, if the residuals come from numerical errors, their lengths have no relation. Using the equivalent of residual length as the criterion to judge whether ghost images exist has excellent numerical performance.

\subsubsection{Slope Detector}

The directions of ghost and real images near critical curves are perpendicular to each other, based on which \twinkle constructs the slope detector. In the case of a binary lens, one point source inside one caustic produces five real images. In contrast, those outside all caustics only produce three real images and two ghost images. Considering that a point source crosses the caustic from the exterior to the interior, a pair of ghost images combine on the critical curve and subsequently diverge into a pair of real images. The complex number argument of images pair difference jump by $\pi/2$.

\begin{figure*}
  \centering
  \hspace{-1.1cm}
  \includegraphics[scale=0.482]{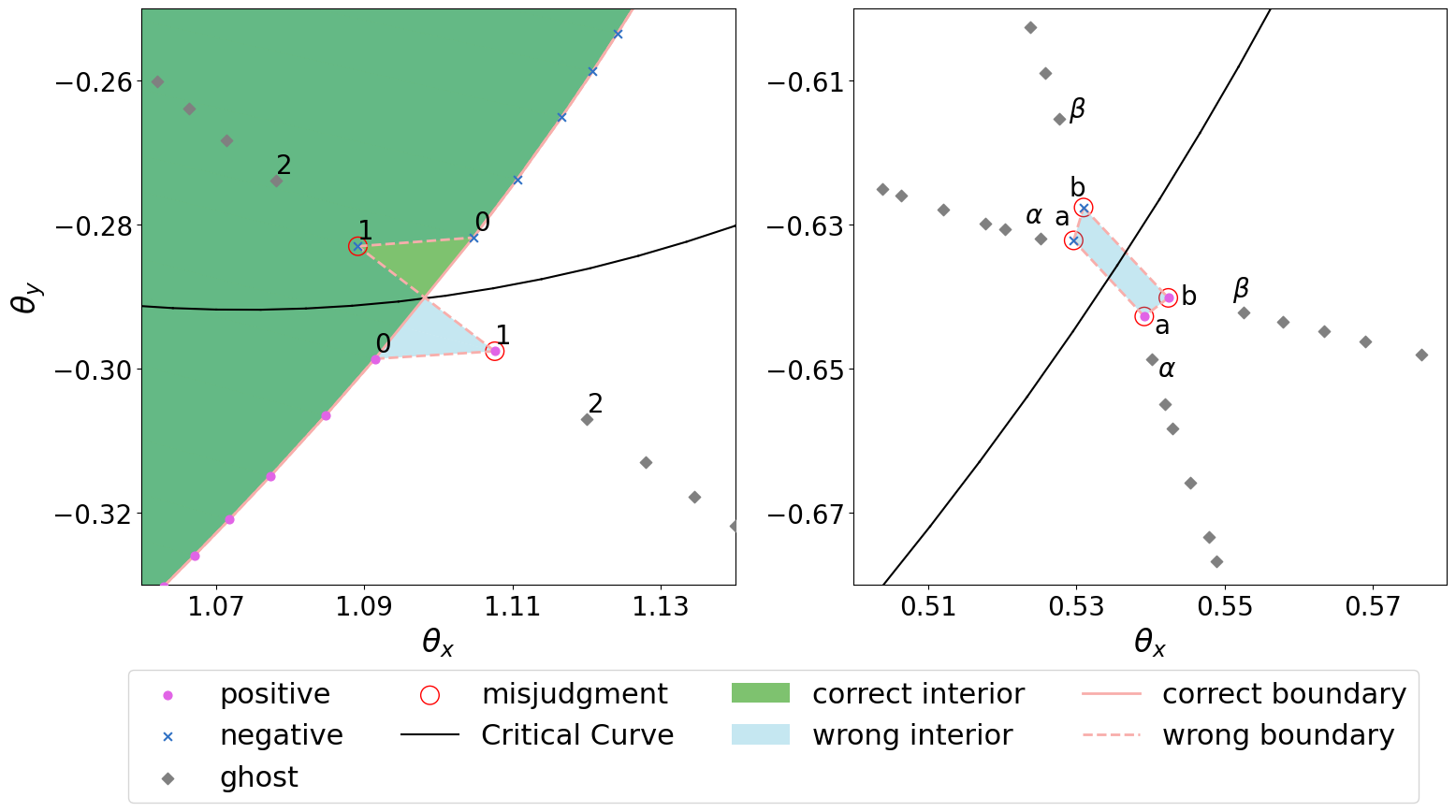}
  \caption{Schematic examples of ghost image pair misjudgment near the critical curve. The magenta dots and the blue crosses are real images with positive and negative parity, while the gray diamonds are ghost images. The black line is the critical curve, on which a pair of real images could transfer into a pair of ghost images. Image pairs marked with the same text come from the same point sources. 
  In the left panel, pair $1$ are ghost images misjudged as real images, forming a wrong boundary of the extended image. The correct boundary near the critical curve is the connection of pair $0$ images, as shown by the pink solid line. In the right panel, pair $a$ and $b$ are misjudged as real images, forming a fake piece of the image and may cause spikes in the light curves.
  }
  \label{fig:phys}
\end{figure*}

The angle of $\pi/2$ could be explained by the local expansion \citep{1992grle.book.....S, 2010MNRAS.408.2188B}. With some expanding coefficients $\lambda$, $a$, and $b$, the difference of the image pairs near the critical curve takes the following form: 
\begin{equation}\label{Delta_z}
    \boldsymbol{z}_{+} - \boldsymbol{z}_{-} = \frac{2\sqrt{2a \mathrm{Re}(\zeta)\lambda^2 + (b^2-ac)\mathrm{Im}(\zeta)^2}}{a\lambda} .
\end{equation}
When a point source, $\zeta$, moves from the interior to the caustic exterior, the value under the square root in Equation~\eqref{Delta_z} continuously drops across $0$. A negative value under the square root generates an $\mathrm{i}$ factor, which causes a rotation of $\pi/2$.

Figure \ref{fig:phys} shows two typical situations the slope detector tries to avoid. The scattering points are produced by points on the boundary of the source, while different marks represent different image types. Image pairs with the same text are generated from the same point sources, and the pairs marked by open red circles are ghost image pairs that are misjudged as real. In the left panel, the source boundary crosses the caustic. Therefore, the pair $0$ should be connected across the critical curve to form the image boundary. However, the ghost pair $1$ is misjudged as a real pair and disturbs the boundary into the dashed line, and the wrong image interior gives out an unreliable image area. Besides, the misjudgment also causes great trouble in the adaptive sampling part since the program believes cross-caustic happens between pairs $1$ and $2$, instead of pairs $0$ and $1$. Consequently, many new sampling points are set between pairs $1$ and $2$, which is useless for detecting information closer to the critical curve between pairs $0$ and $1$.

The right panel in Figure \ref{fig:phys} shows the case when the source is out of the caustic but the boundary is close to it. Therefore, two tracks of ghost images are close to the critical curve and may produce fake image pieces. The image pair $a$ and $b$ are ghost images misjudged as real images, forming a quadrilateral that should not exist. This mistake adds an additional area to the total area, resulting in a wrong magnification. Such a misjudgment usually causes spikes in simulated light curves.

The slope detector checks whether the corresponding image pairs are perpendicular enough. As shown in the left panel of Figure \ref{fig:phys}, the real images and ghost images are perpendicular to each other near the critical curve. When the slope detector finds that the perpendicular condition is not satisfied, nearby image pairs are checked until the slope jump position is found or reaches the end of the image piece. In the left panel, pairs $1$ and $2$ are not perpendicular, but pairs $0$ and $1$ are. Therefore, pair $1$ shall be rejudged as ghost images, pair $0$ will be connected, and then the problem will be fixed. In the right panel, neither pair $a$ and $\alpha$ nor pair $b$ and $\beta$ satisfy the perpendicular condition. As a result, the whole area piece shall be deleted, and the corresponding images are repaired.

\subsection{Error Estimate and Adaptive Sampling}

Error estimation is used to judge when to stop iteration, while adaptive sampling is crucial to increasing iteration efficiency. The calculation finishes when the area error is better than the preset tolerance. The adaptive sampling technique sets more sampling points in positions generating higher error.

\twinkle adopts the parabolic correction forms in \cite{2010MNRAS.408.2188B} and \cite{VBBL4}. For each parabolic correction $\mathrm{d}A^{(p)}_i$, there are four kinds of error: $E_{1,i}, E_{2,i}, E_{3,i}, E_{4_i}$. In usual cases, 
\begin{equation}\label{eq:area_normal}
\begin{aligned}
    E_{1,i} &= \frac{1}{48} \left|(\boldsymbol{z}' \wedge \boldsymbol{z}'')|_{\theta_i} - (\boldsymbol{z}' \wedge \boldsymbol{z}'')|_{\theta_{i+1}} \right| \Delta \theta^3 , \\
    E_{2,i} &= \frac{3}{2}\left|\mathrm{d}A^{(p)} \left(\frac{|z_i - z_{i+1}|^2}{\Delta \theta^2 |z_i' \cdot z_{i+1}'|} -1 \right) \right| , \\
    E_{3,i} &= \frac{1}{10}|\mathrm{d}A^{(p)}_i|\Delta \theta^2 , \\
    E_{4,i} &= |\mathrm{d}A^{(p_1)}_i - \mathrm{d}A^{(p_2)}_i|.
\end{aligned}
\end{equation}

In cross-caustic cases, 
\begin{equation}\label{eq:area_cross}
\begin{aligned}
    E_{1,i} = &\frac{1}{48} \left|(\boldsymbol{z}' \wedge \boldsymbol{z}'')|_{+,i} + (\boldsymbol{z}' \wedge \boldsymbol{z}'')|_{-,i} \right| \widetilde{\Delta \theta}^3 ,  \\
    E_{2,i} = &\frac{3}{2}\left| (z_{+,i} - z_{-,i})^2 \cdot (z_{+,i} - z_{-,i})^2 \right.\\
    &\left. \mp 2|z_{x,i} - z_{-,i}|\sqrt{|z'_{+,i} \cdot z'_{-,i}|}  \right| \widetilde{\Delta \theta}, \\
    E_{3,i} = &\frac{1}{10}|\mathrm{d}A^{(p)}_i|\Delta \theta^2 .
\end{aligned}
\end{equation}
In the cross-caustic form of $E_{2, i}$, $-$ is adopted when two new real images are created along the positive direction of $\theta$,  while the $+$ sign is just the opposite.

\begin{figure}
\centering
\hspace{-0.7cm}
\includegraphics[scale=0.5]{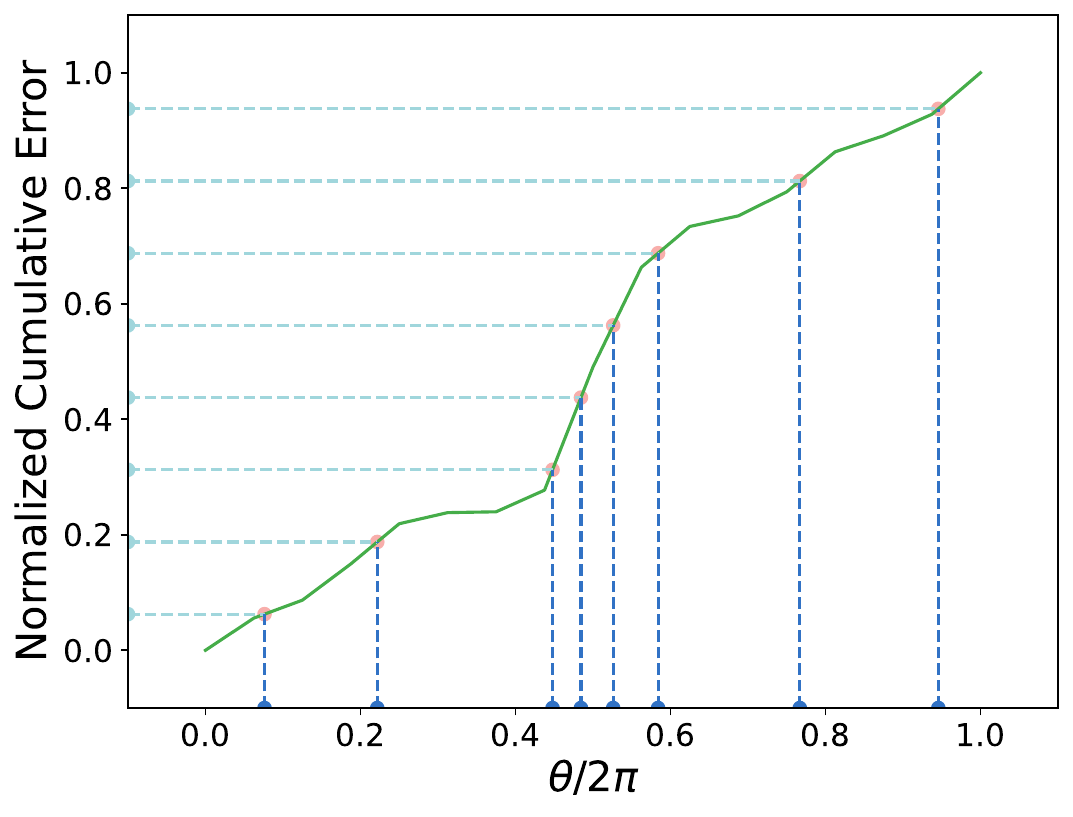}
\caption{The inverse transform method generates eight new samples. $\theta$ is the position angle of the sampling point source. The green line is the cumulative distribution function induced by the area error. The cyan quantiles distribute uniformly on the vertical axis, mapping to $\theta$ positions by finding the points on the cumulative distribution curve. Since cumulative errors rise sharply in the positions with high error, more sampling points are attracted there.}
\label{fig:inverse-transform}
\end{figure}

Since different positions have different errors, more sampling points are required to be located in high-error positions. The adaptive sampling strategy is using the error in each interval $(\theta_i, \theta_{i+1})$ as the sampling weight. With the sampling point density proportional to the area error, \twinkle uses the inverse transform method to generate new sampling points. As shown in Figure \ref{fig:inverse-transform}, the green solid line is the normalized cumulative error. For each $\theta$, the green line is the area error between $[ 0, \theta )$ divided by the total area error. The adaptive sampling method of \twinkle chooses a set of quantiles ($64$ quantiles in default), finds the points whose vertical coordinates are equal to the quantiles on the normalized cumulative error line with linear interpolation, then the horizontal coordinates of which are the new sampling $\theta$.
Positions with higher error increase sharply in cumulative error. Therefore, the adaptive sampling strategy attracts more sampling points in the high-error region.

A useful trick is, since we set the tolerance before, we subtract each error by the average tolerance. In other words, if the error between $(\theta_i,\theta_{i+1})$ is already lower than expected, no more points needed to be added here. This trick is helpful to accelerate convergence.

\section{GPU Implementation}
\label{sec:GPU}

Due to the hardware design of GPUs, program design has many details to consider to ensure maximum efficiency in program execution. In GPU programming, there are three levels of computation hierarchy: thread, block, and grid. A thread is the smallest computation unit; each thread can request its independent memory space. Computations of different threads are parallel to each other and generally do not interfere with one another. A block is a collection of multiple threads, and due to hardware architecture constraints (warp size), the number of threads in a single block is typically a multiple of $32$. Threads within the same block can set synchronization points and perform some low-bandwidth data exchanges. However, due to limited hardware resources, the size of a block is also constrained, and more threads need to be allocated to different blocks to form a grid. Similar to the computation hierarchy, GPU memory is divided into multiple levels, primarily including the fastest-access registers, the shared memory visible within a block, and the largest-capacity global memory visible to all but with slower read speeds.

The overall iterative computation process of \twinkle is roughly as follows. In each iteration, $64$ sampling point sources are selected from the contour; the position of the point images is calculated, and the error is determined; the area of each connected image pair is summed to obtain the magnification and error for the current iteration step; if the error is less than the tolerance, the results are outputted; otherwise, $64$ new sampling points are allocated for the next iteration based on the current error information. Different tasks are assigned to different levels of computation to ensure higher computational efficiency.

\twinkle employs a single thread to calculate a single point source, utilizing the fastest-accessible registers for intermediate variables during the computation process. Results that need to be reused are stored back in the global memory. The computational content for a thread includes the position of the five images, the parity of each image, the use of the residual detector to discern real from ghost images, the connection of adjacent images, and the calculation of the corresponding area. A noteworthy detail is that the computation of $z' \wedge z''$ in  Equation~\ref{eq:area_normal} (see also Equation~(24) in \cite{2010MNRAS.408.2188B}) is computationally intensive and requires multiple reuses; hence, it is also stored in the global memory for subsequent access. In each iteration, calculations that only involve a single sampling point (such as solving for the positions of the five images) are performed only for the newly added sources. In contrast, calculations that involve adjacent sources (such as the connection of adjacent points) necessitate modifications to both the current iteration's source and the results of the adjacent source in the shared memory.

The computation for each extended source is managed by a single block, where the most critical steps involve the allocation of contour sampling sources and the summation of area. Typically, in each iteration, there are $64$ sampling points (the number is adjustable) on an extended source that are computed simultaneously, and these threads are allocated within the same block. Due to the hardware design of GPUs, the summation process within a block requires the use of shared memory for parallel reduction summation. Given the limited size of shared memory, the number of cumulative errors during the adaptive step can become a bottleneck, limiting the maximum number of sampling points. 
For a block with $64\ \mathrm{kB}$ shared memory, the current upper limit for sample points on the contour is $4096$ per source.
This limit is reached only if exceedingly small tolerance values for cross-caustic situations are imposed (for example, tolerance $\ll 10^{-6}$ for mass ratio $q \ll 10^{-6}$).
In summing up the areas of the images, only the initial few iterations will sum the areas for all sampling points. At the same time, subsequent steps will only calculate the changes for the newly added sampling points to ensure greater computational efficiency.
If all $4096$ points are used up, \twinkle will return the magnification computed with $4096$ points and the estimated error.

Multiple blocks are allocated on the GPU to compute multiple extended sources concurrently since many data points are on the light curve. Since the computation time for magnifications varies across different locations, a more sophisticated task dispatch is required to prevent idle computation resources when only a few extended sources undergo multiple iterations while others have completed their calculations. \twinkle employs a method based on multiple GPU streams for allocating computational tasks. The streams operate independently of one another, with no hierarchical relationships. Consequently, when idle computational resources are present on the GPU, pending tasks are automatically allocated to the available streaming multiprocessors.

In the \twinkle framework, atomic operations are utilized for several critical functions, including recording positions across caustics, marking the adjacent point sources for newly added sample points. Although these atomic operations are not performance bottlenecks, it is necessary to ensure that conflicts do not arise when multiple threads perform operations on the same memory locations simultaneously, which is particularly important when using different types of GPUs.

\twinkle also adopts a compatibility layer from the numerical simulation code Kratos (L. Wang 2025, in prep.) to insulate the implementation of algorithms from the actual programming models. This approach enables the application of \twinkle on multiple heterogeneous architectures and programming models, including, e.g., CUDA for NVIDIA GPUs and HIP for AMD GPUs and DCUs, etc.

\section{Computational Performance}
\label{sec:speed}

With previous modifications, especially the coefficient generator and ghost image detector, \twinkle achieves better calculation speed and robustness. Since \twinkle is designed to search for exoplanets,
performance tests mainly focus on the positions that can highlight the perturbation effects of planets. In simple terms, it is around the positions near the planetary caustic.
To demonstrate the improvement of performance of \texttt{Twinkle}, the computational speeds are compared with \vbbl \citep[C++ version]{2018MNRAS.479.5157B} for identical cases as examples.
Unless specifically noted, we conduct the speed and accuracy tests on the computer platform equipped with both GPU (NVIDIA RTX 4090, clock frequency $2.2 {\rm GHz}$) and CPU (AMD EPYC 7H12, clock frequency $2.6 {\rm GHz}$). For fair comparisons, each test utilizes one CPU core when using \texttt{VBBL}, or one GPU when \twinkle is used. The \vbbl function used in this article is \texttt{BinaryMag2}. Both \twinkle and \vbbl have enabled the \texttt{-O3} optimization option for compilation.

\subsection{Speed Comparison on the Source Plane}

\begin{figure*}
  \centering
  \vspace{-1.7cm}
  \includegraphics[width=15cm]{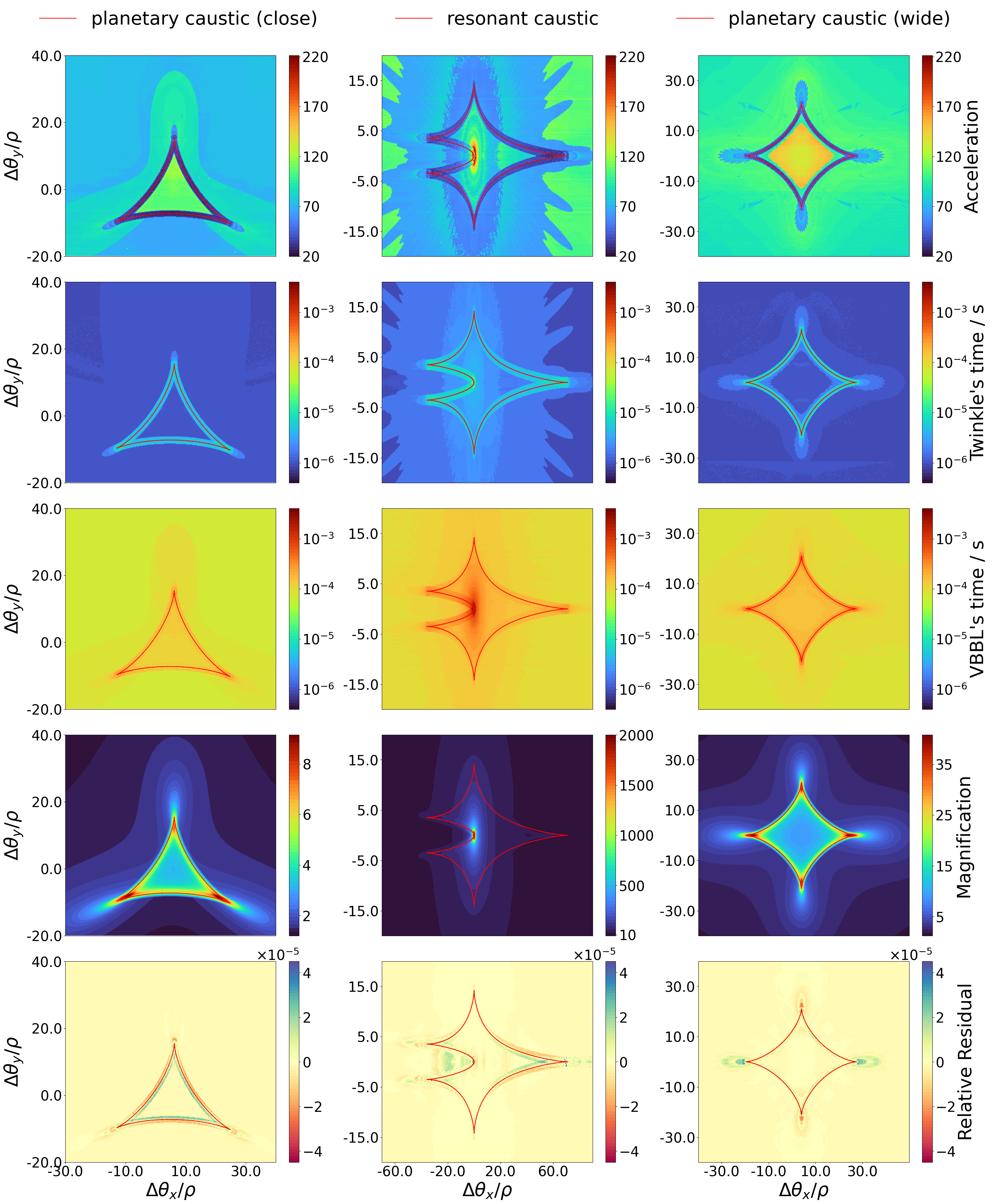}

  \caption{Calculation performance of \twinkle compared with \texttt{VBBL}. All the origin points are moved to the position of planetary caustic (or the resonant caustic for $s=1$). The left column takes $s=0.5, q=10^{-4}, \rho = 5\times 10^{-5}, \Delta \theta_x = (\theta_x + 1.5), \Delta \theta_y= (\theta_y - 0.035)$.
  The middle column takes $s=1, q=10^{-4}, \rho = 1\times 10^{-3}, \Delta \theta_x = \theta_x, \Delta \theta_y= \theta_y$.
  The right column takes $s=3, q=10^{-4}, \rho = 1\times 10^{-4}, \Delta \theta_x = (\theta_x -2.666), \Delta \theta_y= \theta_y$. The red solid line is the caustics.
  The performance for specific extended sources is shown by the color at the source center location. The origin point is moved to the caustics, meaning $\Delta \theta_x$ and $\Delta \theta_y$. Each row except the fourth row takes the same color map. Additionally, the calculation times for \twinkle and \vbbl in the second and third rows share the same logarithmic color mapping to demonstrate the calculation features.
  }
  \label{speed_comp}
  \title{\twinkle Performance}
\end{figure*}

Different source locations require different calculation abilities. Figure \ref{speed_comp} shows the computational performance of \twinkle
compared with \texttt{VBBL}. With a tolerance of $10^{-4}$ relative error and mass ratio $q = 10^{-4}$, we set three different $s$ values, representing three typical types of caustics. The number of sampled sources is $480 \times 340$, while the source radius is selected to match the scale of the planetary caustic. It is noted that the source radius in the first column is smaller than the typical scale for source stars in the Galactic bulge ($10^{-4}\lesssim \rho \lesssim 10^{-3}$ ). However, similar selections of parameters also occur in actual calculations using contour integration, in which the spatial scales of contours near the center are usually $1$ order of magnitude smaller than the source radius.

As one can observe from the bottom row of Figure~\ref{speed_comp}, the robustness of \twinkle is verified down to the mass ratio $q = 10^{-4}$. The bottom row calculates the relative difference in magnification obtained by \twinkle and \texttt{VBBL}. 
In \twinkle's computation, an error estimate is made at each iteration. The iteration stops and outputs results when the error is less than the preset tolerance.
The difference between \twinkle and \vbbl tends to be bigger near the caustic, especially around cusps. However, all the results are still within the preset relative tolerance of $10^{-4}$ successfully.

The speed of \twinkle on one GPU is equivalent to $\sim 10^2$ times the \vbbl on one CPU core at each point involved in the computation, as illustrated in the first row of Figure~\ref{speed_comp}. 
The features in the acceleration row have diverse origins. The second and third rows share the same logarithmic color mapping and demonstrate the computational speed of \twinkle and \texttt{VBBL}. Solving for the position of the source across the caustic line is particularly challenging, requiring more computational time for both \twinkle and \texttt{VBBL}. However, when the source is located inside or outside the caustic line, \twinkle exhibits a more significant acceleration in computation.

Considering the features of GPU computation, \twinkle has implemented special optimizations for task dispatch. For a series of extended sources requiring solving, each iteration marks whether their calculation is already finished.
Uncompleted tasks are recorded in a list, and only these sources are allocated to the GPU for computation in the subsequent iteration. This dispatch strategy prevents the computation from being blocked at high-cost cross-caustic positions, improving overall computational efficiency.

Note that two cases have \texttt{VBBL}'s results that exceed the designated error tolerance, and these two outliers are excluded in Figure~\ref{speed_comp}. Both cases have resonant caustics (middle panel), and the problematic source positions result in high magnifications of $\sim10^3$, which occur rarely. With only two outliers outside of $480 \times 340$, the numerical issue is minor for $q = 10^{-4}$. As demonstrated in the next section (\S~\ref{sec:lowq}), the numerical problems occur more frequently for smaller mass ratios with \texttt{VBBL}.

\subsection{Lower Mass Ratio}
\label{sec:lowq}

\begin{figure*}
\centering
\includegraphics[width=\textwidth]{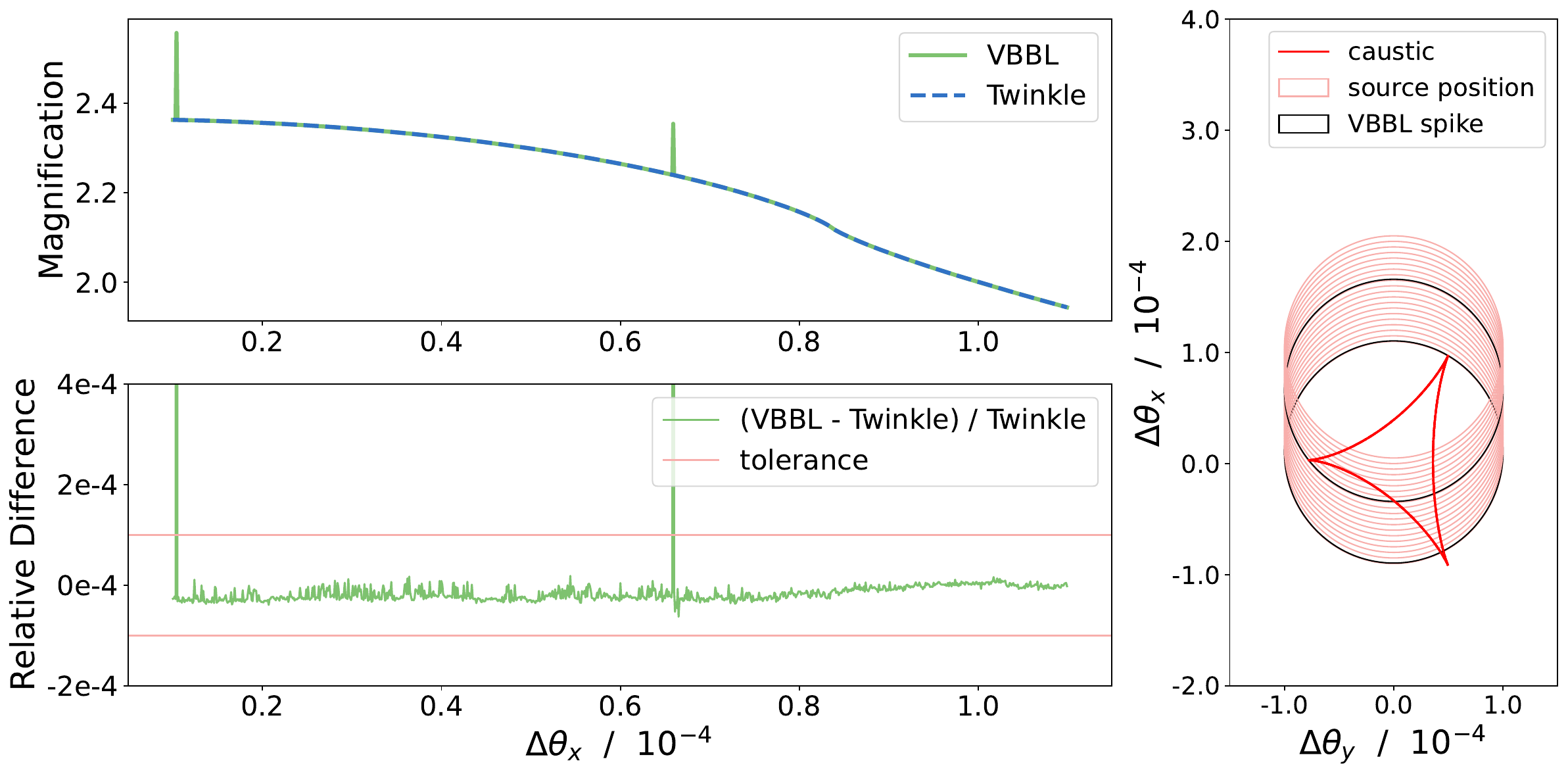}

\caption{An example of \twinkle robustness. The upper left panel shows the magnification calculated by \twinkle and \texttt{VBBL}, and the lower left panel shows the relative difference, with the horizontal pink lines representing tolerance and the horizontal axis representing the coordinate of source centers. The right panel is the relative position of the source and the planetary caustic, in which the red line is a planetary caustic, and the circles are extended source positions corresponding to data points in the left panels. The origin point is moved to the position of the planetary caustic, where $\Delta \theta_x = \theta + 1.5000$, $\Delta \theta_y = \theta_y + 0.0035$. The spikes in the left panel correspond to the black circles in the right panel, whose boundaries are near the cusp of the planetary caustic. The upper and lower black circles correspond to the right and left spikes, respectively.}

\label{correcter}
\end{figure*}

Numerical challenges increase when the planet-star mass ratio $q$ becomes smaller, where \twinkle adopts multiple optimizations to confront these challenges for more reliable performance. Figure \ref{correcter} illustrates the light curve for a $q = 10^{-6}$ example,  which includes cases where the boundary of the source is very close to the caustic, representing the most numerically challenging positions.
The lens parameters are $s=0.5$, $q=10^{-6}$, and the source radius is $\rho = 10^{-4}$. The $\theta_y$ coordinate is fixed at $\theta_y = -0.0035$. The upper left panel shows the magnification calculated by \twinkle and \texttt{VBBL}, revealing an anomaly in \vbbl's results. The lower left panel displays the relative difference between the two algorithms' results, with the two horizontal pink lines representing the preset tolerance of the program ($10^{-4}$). The right panel illustrates the relative position of the source and the planetary caustic.

One can notice two abnormal "spikes" appearing in the magnification results of \vbbl at $\theta_x = -1.49998960$ and $\theta_x = -1.49993410$, corresponding to the source position marked with black in the right panel of Figure~\ref{correcter}. At this position, points on the boundary are close to the cusp, introducing significant numerical errors and even incorrect connections. Once incorrect connections occur, the magnification differs significantly from the true value, and the subsequent fitting procedures utilizing the contour integration algorithm can hardly converge. 
The computation speed also suffers from this divergent behavior, because  the adaptive sampling step cannot locate extra sample points to the correct positions either.

\subsection{Exploring the Parameter Space}\label{sec:sq}

\begin{figure*}
\centering
\hspace{-0.3cm}
\includegraphics[width=\textwidth]{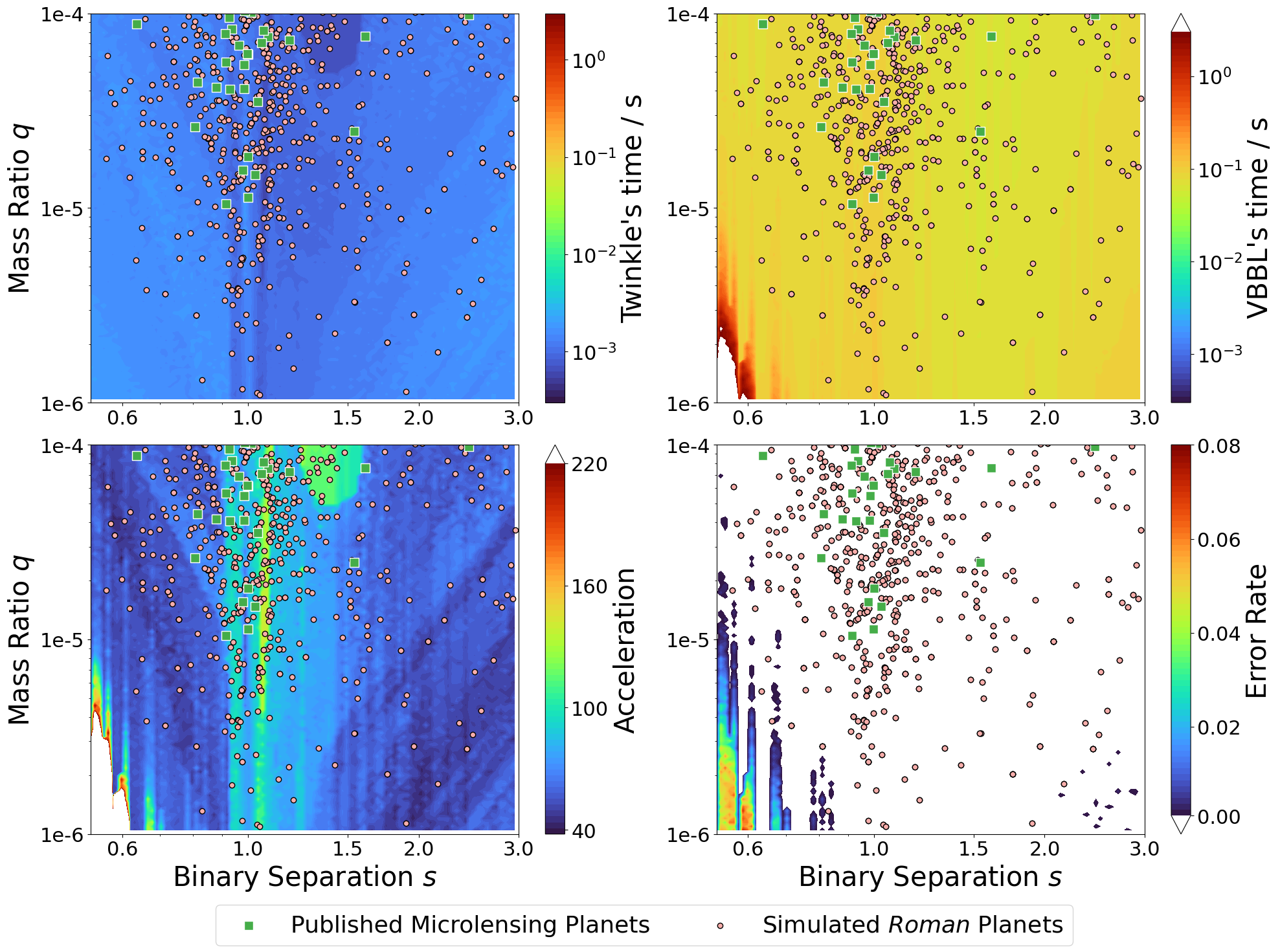}

\caption{Calculation performance in parameter space. The upper row shares the same logarithmic color mapping, in which the blank region marks where \vbbl takes abnormally long calculation time. The bottom left panel is acceleration defined as the time ratio of \twinkle and single-thread \texttt{VBBL}. The bottom right panel is the error rate as the proportion of \vbbl wrong results that exceed the tolerance compared to the results from \texttt{Twinkle}. The white region in the bottom right panel means no differences between \twinkle and \vbbl magnification results. The green squares and pink dots are the published microlensing events and the simulated \textit{Roman} detection.}

\label{speed_sq}
\end{figure*}

The parameter space spanned by different $\{s, q\}$ values is explored to verify the accuracy and robustness of \twinkle in Figure~\ref{speed_sq}, which also illustrates the speed performance under various conditions compared to \texttt{VBBL}. 
For each $\{s,q\}$ value, the selection of source positions includes a $30 \times 30$ grid of points that thoroughly covers the planetary or resonant caustic. The extended source has radius $\rho = 10^{-4}$ for the typical radii of stars in the bulge field.
The scattered points are microlensing events, in which the green squares are already-published events mostly discovered by ground-based surveys from the \cite{ml}, while pink points are simulated \textit{Roman} detections from \cite{Penny2019}.

In general, \twinkle keeps a consistent and robust computational performance over a broader range of parameter space.
The upper two panels in Figure~\ref{speed_sq}  sharing the same logarithmic color mapping compare the calculation speed of \twinkle and \texttt{VBBL}, where the blank region marks where \vbbl encounters an abnormally long calculation time. As one can observe, this abnormal region mainly happens in $q\lesssim 10^{-5}$ and $s\sim 0.6$, which is highly relevant in the proposed search of Earth-like planets via microlensing observations. 
The acceleration in the bottom left panel is the ratio of time used by \twinkle and \vbbl for the same task. Usually, the calculation speed of \twinkle is about $100$ times faster than the single-thread \texttt{VBBL}, similar to the number of stream multiprocessors (SMs) on one NVIDIA RTX 4090 GPU. However, the abnormal region is not present in the \twinkle calculation, which is crucial for detecting low-mass exoplanets.

We specifically note that \vbbl fails to yield correct magnifications over the relatively broad range of parameter space concerned here due to multiple algorithmic issues (not to be confused with the numerical errors), especially incorrect connections of points. The bottom right panel in Figure~\ref{speed_sq} shows the error rates of \texttt{VBBL}, defined as the proportion of erroneous points among all sampled points, in which the white region indicates where the magnifications calculated by \twinkle and \vbbl are consistent. Similar to the abnormal region in the acceleration panel, these incorrect convergences mainly happen in $q\lesssim 10^{-5}$ and $s\lesssim 0.7$, which produces great trouble for detecting Earth-like exoplanets. \twinkle can correctly and efficiently solve all magnification within the parameter space. This capability is attributed to \texttt{Twinkle}'s more precise generator for polynomial equation coefficients and more robust ghost image detector. We have conducted additional tests demonstrating that our code can reliably calculate magnifications for binary separations in the range $0.1 \leq s \leq 10$ and mass ratios $q \geq 10^{-9}$ with a relative tolerance of $10^{-4}$. This parameter space likely covers most of the practical applications anticipated in the foreseeable future.

A slight slowdown in the calculation speed can be observed in the region of $s \approx 1$, predominantly due to the inclusion of source positions very close to the host star with high magnification. As indicated in the second column of Figure \ref{speed_comp}, the number of positions that take the longest to compute is limited. These conditions are not commonly encountered during actual observations. Therefore, the region where $s$ is close to $1$ does not result in a significant degradation in performance during actual detection.

The reliability of \twinkle will be valuable for highly sensitive space-based microlensing projects. As Figure~\ref{speed_sq} shows, the known microlensing planets (green squares) discovered from ground-based surveys are hardly affected by the abnormal region.  However, future microlensing projects, such as \textit{Roman}, are expected to explore a much broader parameter space than ground-based surveys. Simulated \textit{Roman} microlensing planets (pink dots; \citealt{Penny2019}) with low $q$ overlap with the abnormal region, where \vbbl takes $10^2$ times longer computation time than usual (see the upper right and lower left panels of Figure~\ref{speed_sq}). Although only a small fraction of the simulated \textit{Roman} planets fall directly into the abnormal region, a larger fraction of events is likely to be affected during the initial parameter space explorations that broadly probe the parameter space.  Additionally, outside the abnormal region, occasional problematic computations can arise, potentially slowing down the Monte Carlo Markov Chain when such occurrences are included. Thus, the stability of \twinkle ensures consistently efficient computations, making it well suited for systematic modeling efforts in \textit{Roman} and similar missions.

\subsection{Scalability}

Computational scaling is used to assess the performance of heterogeneous computing across different scales of calculation tasks, usually categorized into strong scaling and weak scaling. Strong scaling refers to maintaining a constant scale of the calculation task while observing how the computation time changes as more devices are added. In contrast, weak scaling involves increasing the task scale proportionally with the number of devices and examining how the computation time is affected. Due to the special architecture of GPUs, the relationship between computation time and task scale is often nonlinear, especially when communication costs are significant.

\begin{figure}
\centering
\hspace{-1cm}
\includegraphics[scale=0.5]{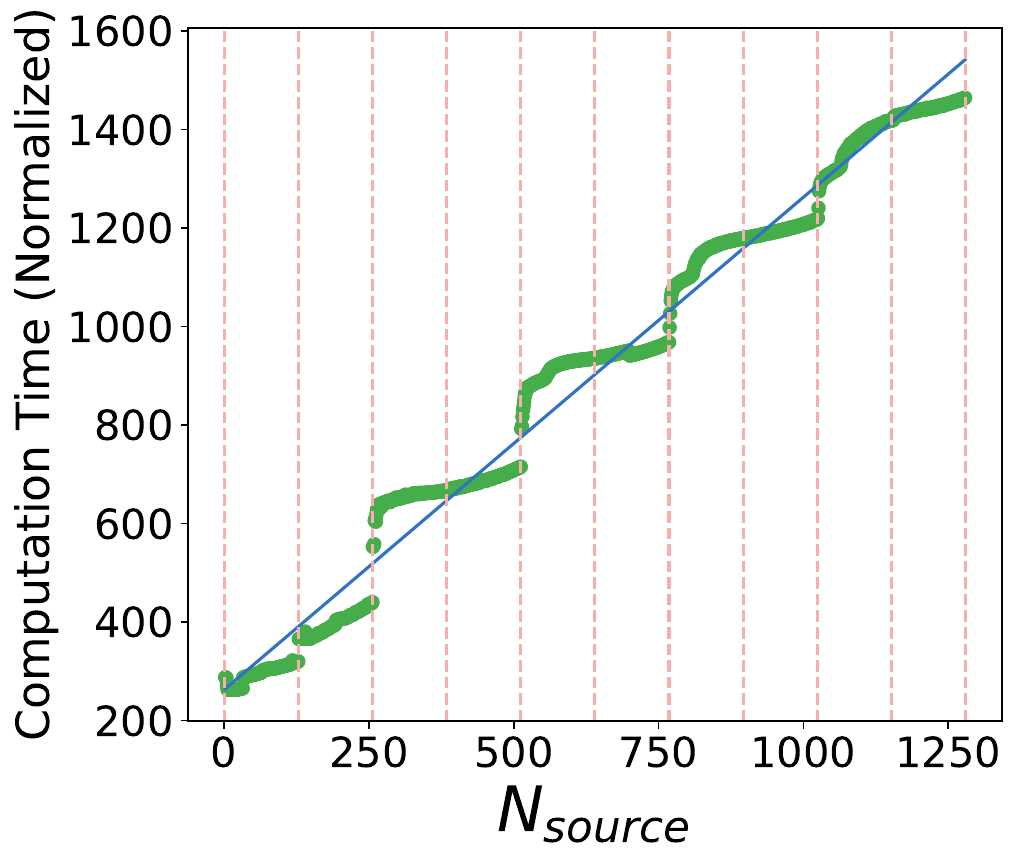}
\caption{Scaling relationship of computations performed by \texttt{Twinkle}. The horizontal axis is the number of sources in tasks, and the vertical axis is the computational time normalized by the linear fitting coefficient. Each source is dispatched to one block. The green data points are the calculation time used for different task scales. The blue solid line is the linear fitting of the task scale and time consumed. The pink vertical lines are multiples of the number of SMs. For the NVIDIA 4090 used in this paper, $N_{{\mathrm{SM}}} = 128$.}
\label{scaling}
\end{figure}

In microlensing modeling, the number of sources required for computation is determined by the observations rather than by the modeler. Therefore, it is essential to test the time taken to compute varying numbers of sources. The scalability test shown in Figure \ref{scaling} is the variation in computational time as the scale of computation increases by \texttt{Twinkle}, which is slightly different from strong scaling and weak scaling. The reason for such a test is that the design of \twinkle handles each extended source in one block and no communication between devices, leading to trivial time variation with the number of devices. 
The horizontal axis represents the scale of the task, that is, the number of extended sources sent to GPU $N_{\mathrm{source}}$. Since \twinkle uses one block for one extended source, $N_{\mathrm{source}}$ equals the number of blocks $N_{{\mathrm{block}}}$. The vertical axis is the normalized computational time, while the normalization coefficient comes from the linear fitting of scale and time. 

With similar time consumed, a scaling relation closer to linear is better. Overall, the \twinkle computational time increases approximately linearly with the computational scale. It experiences some step increases when the extended source number exceeds an integer multiple of $N_{{\mathrm{SM}}}$. $N_{{\mathrm{SM}}}$ is the number of stream SMs, which is a component within a GPU. One GPU typically contains dozens or just more than $100$ SMs. On a larger scale, the computational time also exhibits fluctuations with a periodicity of $2N_{{\mathrm{SM}}}$, originating from the automatic dispatch of blocks on SMs. 
More than one block could be dispatched on the same SM, depending on the occupancy of kernel functions.
The startup time is approximately the computational duration required for $200$ sources. Therefore, it is recommended to have the number of sources in one computation be more than $200$ and slightly less than an integer multiple of $N_{{\mathrm{SM}}}$ for optimal performance.

\subsection{CPU version}

Although \twinkle is optimized for GPUs, our new coefficient generator and ghost image detector are universal and can also improve calculations on CPUs. Therefore, we have also developed a separate C++ version for CPU use\footnote{\href{https://github.com/AsterLight0626/Twinkle_CPU}{https://github.com/AsterLight0626/Twinkle\_CPU}}. This CPU version fully inherits the algorithm of the GPU version, except for replacing parallel computations with loop structures. As a result, the CPU version maintains the same computation robustness as the GPU version.

\begin{figure*}
\centering
\hspace{-0.3cm}
\includegraphics[width=\textwidth]{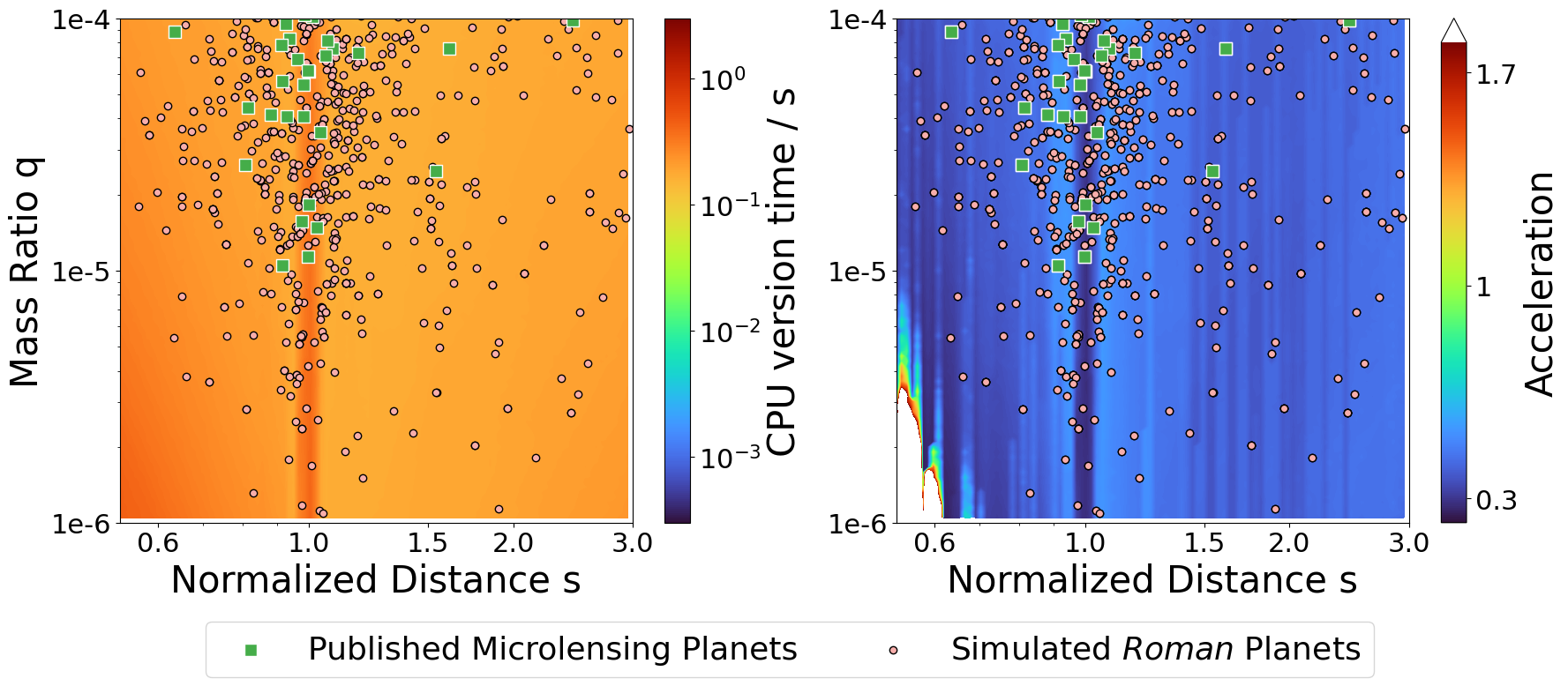}
\caption{CPU version performance in parameter space. The left panel shows the computational time used by the CPU version of \texttt{Twinkle}, with the same color mapping as the upper row of Figure \ref{speed_sq}. The right panel is the acceleration defined as the computation time ratio of the single-thread \texttt{VBBL} and single-thread CPU version of \texttt{Twinkle}. The green squares and pink dots are the published microlensing events and the simulated \textit{Roman} detection.}

\label{fig:cpu}
\end{figure*}

The computational speed of the CPU version is shown in Figure \ref{fig:cpu}, in which the CPU version of \twinkle executes the same tasks as in \S\ref{sec:sq}, with the same CPU (AMD EPYC 7H12). The left panel shows the time used for calculations, sharing the same color mapping as the upper row of Figure \ref{speed_sq}. 
The computational cost is generally similar with \texttt{VBBL}, and it also shares the same slight slowdown near $s \approx 1$ as the GPU version. However, the abnormal region of \texttt{VBBL} in the bottom left corner does not exist for \texttt{Twinkle}-CPU. The right panel is the acceleration defined as the computational time ratio between \vbbl and \texttt{Twinkle}-CPU, showing that in most cases, their computation times are similar, within $\sim 0.5$ to $\sim 1.5$ of \texttt{VBBL}. In the bottom left abnormal region, the time used by \vbbl can be $\sim50$ times longer than \texttt{Twinkle}-CPU.

The performance shown in Figure \ref{fig:cpu} indicates that the CPU version \twinkle keeps the high precision and robustness of the standard GPU version. Although the algorithm is not specially prepared for sequential computing architecture like CPU, it still has a comparable speed to the well-developed \vbbl algorithm. Besides, the CPU version allows more accessible programming for a single event calculation and more convenient cooperation with other code modules, significantly enhancing the versatility of \texttt{Twinkle}.

\section{Summary \& Conclusion}

In this paper, we develop a code to calculate the magnification of a binary-lens microlensing system: \texttt{Twinkle}. It employs the contour integration method, similar to \texttt{VBBL}, while it is optimized for enhancing computational performance on GPUs.

\twinkle incorporates high-precision methods for generating the coefficients for the binary-lens equation (see \S\ref{sec:coeff_generator}, Equation~\ref{eq:method-transformed}). This approach improves the precision of coefficient calculations, reducing the numerical errors introduced by cross-scale arithmetic operations substantially. Such an approach of generating coefficients can be crucial, especially for modeling planetary lens systems with very small mass ratios (e.g., $q\lesssim 10^{-5}$), which is highly relevant in the search for Earth-mass and sub-Earth-mass planets via existing and planned microlensing projects. 
Thanks to the high-precision coefficients, \twinkle is more reliable at forming image contours near the critical curve than previous methods without corrections for numerical errors in cross-scale subtraction.
The coefficient generation process modification is also sufficiently straightforward, which can be easily implemented to other codes that involve the binary-lens equation and especially contour integration.

In addition, \twinkle also develops a new ghost image detection method, which includes two components: a residual detector and a slope detector.
The residual detector is applicable to ghost images of point sources. For the residual detector, we prove that the sum of the residuals of the binary-lens polynomial equation equals $0$
(see also \S\ref{sec:ghost_detector}, Equation~\ref{sum_r} and \ref{sum_r2}),
ensuring that the residual magnitudes are equal when there are two ghost images.
The slope detector is based on the abrupt change in slope when the source boundary crosses a caustic, as depicted in Figure \ref{fig:phys}, giving a more precise refinement of the residual-based ghost image detector results with the information of adjacent point sources near the caustic. 
By avoiding the wrong connections of real images near the critical curve, which could produce spurious magnifications, the ghost detector can guarantee reliable magnification results.

With these improvements, \twinkle achieves a significantly expanded parameter space to produce robust magnifications. The results presented in Figure \ref{speed_sq} demonstrate that \twinkle can efficiently calculate the correct magnification at any location within the range of $0.5 < s < 3$ and for $q > 10^{-6}$, which is particularly beneficial for microlensing detection of Earth-mass and smaller exoplanets.

Finally, we demonstrate that \twinkle on a single 4090 GPU can achieve computational speeds about $10^2$ times faster than a single-threaded CPU, and \twinkle also produces more reliable magnification. We expect that \twinkle will significantly contribute to tackling the modeling challenges of microlensing exoplanet search projects in the future.

\section*{Acknowledgements}

We thank the reviewer for helpful suggestions and Zexuan Wu for checking the computation performance. This work is supported by the National Natural Science Foundation of China (Grant No. 12133005) and the science research grants from the China Manned Space Project with No. CMS-CSST-2021-B12. S.D. acknowledges the New Cornerstone Science Foundation through the XPLORER PRIZE.



\bibliographystyle{aasjournal}
\bibliography{twinkle} 

\begin{thebibliography}{}
\expandafter\ifx\csname natexlab\endcsname\relax\def\natexlab#1{#1}\fi
\providecommand{\url}[1]{\href{#1}{#1}}

\bibitem[{{Bennett}(2010)}]{2010ApJ...716.1408B}
{Bennett}, D.~P. 2010, \apj, 716, 1408

\bibitem[{{Bond} {et~al.}(2001){Bond}, {Abe}, {Dodd}, {Hearnshaw}, {Honda}, {Jugaku}, {Kilmartin}, {Marles}, {Masuda}, {Matsubara}, {Muraki}, {Nakamura}, {Nankivell}, {Noda}, {Noguchi}, {Ohnishi}, {Rattenbury}, {Reid}, {Saito}, {Sato}, {Sekiguchi}, {Skuljan}, {Sullivan}, {Sumi}, {Takeuti}, {Watase}, {Wilkinson}, {Yamada}, {Yanagisawa}, \& {Yock}}]{MOA}
{Bond}, I.~A., {Abe}, F., {Dodd}, R.~J., {et~al.} 2001, \mnras, 327, 868

\bibitem[{{Bozza}(2010)}]{2010MNRAS.408.2188B}
{Bozza}, V. 2010, \mnras, 408, 2188

\bibitem[{{Bozza} {et~al.}(2018){Bozza}, {Bachelet}, {Bartoli{\'c}}, {Heintz}, {Hoag}, \& {Hundertmark}}]{2018MNRAS.479.5157B}
{Bozza}, V., {Bachelet}, E., {Bartoli{\'c}}, F., {et~al.} 2018, \mnras, 479, 5157

\bibitem[{{Bozza} {et~al.}(2024){Bozza}, {Saggese}, {Covone}, {Rota}, \& {Zhang}}]{VBBL4}
{Bozza}, V., {Saggese}, V., {Covone}, G., {Rota}, P., \& {Zhang}, J. 2024, arXiv e-prints, arXiv:2410.13660

\bibitem[{{Cassan}(2017)}]{Cassan2017}
{Cassan}, A. 2017, \mnras, 468, 3993

\bibitem[{{Dominik}(1995)}]{1995A&AS..109..597D}
{Dominik}, M. 1995, \aaps, 109, 597

\bibitem[{{Dominik}(1998)}]{1998A&A...333L..79D}
---. 1998, \aap, 333, L79

\bibitem[{{Dominik}(2007)}]{2007MNRAS.377.1679D}
---. 2007, \mnras, 377, 1679

\bibitem[{{Dong} {et~al.}(2006){Dong}, {DePoy}, {Gaudi}, {Gould}, {Han}, {Park}, {Pogge}, {MuFun Collaboration}, {Udalski}, {Szewczyk}, {Kubiak}, {Szyma{\'n}ski}, {Pietrzy{\'n}ski}, {Soszy{\'n}ski}, {Wyrzykowski}, {{\.Z}ebru{\'n}}, \& {OGLE Collaboration}}]{2006ApJ...642..842D}
{Dong}, S., {DePoy}, D.~L., {Gaudi}, B.~S., {et~al.} 2006, \apj, 642, 842

\bibitem[{{Dong} {et~al.}(2009){Dong}, {Bond}, {Gould}, {Koz{\l}owski}, {Miyake}, {Gaudi}, {Bennett}, {Abe}, {Gilmore}, {Fukui}, {Furusawa}, {Hearnshaw}, {Itow}, {Kamiya}, {Kilmartin}, {Korpela}, {Lin}, {Ling}, {Masuda}, {Matsubara}, {Muraki}, {Nagaya}, {Ohnishi}, {Okumura}, {Perrott}, {Rattenbury}, {Saito}, {Sako}, {Sato}, {Skuljan}, {Sullivan}, {Sumi}, {Sweatman}, {Tristram}, {Yock}, {MOA Collaboration}, {Bolt}, {Christie}, {DePoy}, {Han}, {Janczak}, {Lee}, {Mallia}, {McCormick}, {Monard}, {Maury}, {Natusch}, {Park}, {Pogge}, {Santallo}, {Stanek}, {{\ensuremath{\mu}}FUN Collaboration}, {Udalski}, {Kubiak}, {Szyma{\'n}ski}, {Pietrzy{\'n}ski}, {Soszy{\'n}ski}, {Szewczyk}, {Wyrzykowski}, {Ulaczyk}, \& {OGLE Collaboration}}]{2009ApJ...698.1826D}
{Dong}, S., {Bond}, I.~A., {Gould}, A., {et~al.} 2009, \apj, 698, 1826

\bibitem[{{Gaudi}(2012)}]{2012ARA&A..50..411G}
{Gaudi}, B.~S. 2012, \araa, 50, 411

\bibitem[{{Gould}(1994)}]{Gould1994}
{Gould}, A. 1994, \apjl, 421, L75

\bibitem[{{Gould}(2008)}]{Gould2008}
---. 2008, \apj, 681, 1593

\bibitem[{{Gould} \& {Gaucherel}(1997)}]{1997ApJ...477..580G}
{Gould}, A., \& {Gaucherel}, C. 1997, \apj, 477, 580

\bibitem[{{Gould} \& {Loeb}(1992)}]{1992ApJ...396..104G}
{Gould}, A., \& {Loeb}, A. 1992, \apj, 396, 104

\bibitem[{{Han}(2006)}]{2006ApJ...638.1080H}
{Han}, C. 2006, \apj, 638, 1080

\bibitem[{{Kayser} {et~al.}(1986){Kayser}, {Refsdal}, \& {Stabell}}]{1986A&A...166...36K}
{Kayser}, R., {Refsdal}, S., \& {Stabell}, R. 1986, \aap, 166, 36

\bibitem[{{Kim} {et~al.}(2016){Kim}, {Lee}, {Park}, {Kim}, {Cha}, {Lee}, {Han}, {Chun}, \& {Yuk}}]{Kim2016}
{Kim}, S.-L., {Lee}, C.-U., {Park}, B.-G., {et~al.} 2016, JKAS, 49, 37

\bibitem[{{Kuang} {et~al.}(2021){Kuang}, {Mao}, {Wang}, {Zang}, \& {Long}}]{2021MNRAS.503.6143K}
{Kuang}, R., {Mao}, S., {Wang}, T., {Zang}, W., \& {Long}, R.~J. 2021, \mnras, 503, 6143

\bibitem[{{Mao} \& {Paczynski}(1991)}]{1991ApJ...374L..37M}
{Mao}, S., \& {Paczynski}, B. 1991, \apjl, 374, L37

\bibitem[{{Misner} {et~al.}(1973){Misner}, {Thorne}, \& {Wheeler}}]{Misner1973}
{Misner}, C.~W., {Thorne}, K.~S., \& {Wheeler}, J.~A. 1973, {Gravitation} (Princeton University Press)

\bibitem[{{Mroz} \& {Poleski}(2023)}]{2023arXiv231007502M}
{Mroz}, P., \& {Poleski}, R. 2023, arXiv e-prints, arXiv:2310.07502

\bibitem[{{NASA Exoplanet Archive}(2024)}]{ml}
{NASA Exoplanet Archive}. 2024, Microlensing Planets Table, vVersion: 2024-10-22,  NExScI-Caltech/IPAC, doi:10.26133/NEA38.
\newblock \url{https://catcopy.ipac.caltech.edu/dois/doi.php?id=10.26133/NEA38}

\bibitem[{{Nemiroff} \& {Wickramasinghe}(1994)}]{NemiroffWickramasinghe1994}
{Nemiroff}, R.~J., \& {Wickramasinghe}, W.~A.~D.~T. 1994, \apjl, 424, L21

\bibitem[{{Pejcha} \& {Heyrovsk{\'y}}(2009)}]{Pejcha2009}
{Pejcha}, O., \& {Heyrovsk{\'y}}, D. 2009, \apj, 690, 1772

\bibitem[{{Penny} {et~al.}(2019){Penny}, {Gaudi}, {Kerins}, {Rattenbury}, {Mao}, {Robin}, \& {Calchi Novati}}]{Penny2019}
{Penny}, M.~T., {Gaudi}, B.~S., {Kerins}, E., {et~al.} 2019, \apjs, 241, 3

\bibitem[{{Rhie}(2001)}]{2001astro.ph..3463R}
{Rhie}, S.~H. 2001, arXiv e-prints, astro

\bibitem[{{Schneider} {et~al.}(1992){Schneider}, {Ehlers}, \& {Falco}}]{1992grle.book.....S}
{Schneider}, P., {Ehlers}, J., \& {Falco}, E.~E. 1992, {Gravitational Lenses} (Springer Berlin, Heidelberg), doi:10.1007/978-3-662-03758-4

\bibitem[{{Schneider} \& {Weiss}(1986)}]{1986A&A...164..237S}
{Schneider}, P., \& {Weiss}, A. 1986, \aap, 164, 237

\bibitem[{{Schneider} \& {Weiss}(1987)}]{SchneiderWeiss1986}
---. 1987, \aap, 171, 49

\bibitem[{{Schramm} \& {Kayser}(1987)}]{1987A&A...174..361S}
{Schramm}, T., \& {Kayser}, R. 1987, \aap, 174, 361

\bibitem[{{Skowron} \& {Gould}(2012)}]{2012arXiv1203.1034S}
{Skowron}, J., \& {Gould}, A. 2012, arXiv e-prints, arXiv:1203.1034

\bibitem[{{Udalski} {et~al.}(2015){Udalski}, {Szyma{\'n}ski}, \& {Szyma{\'n}ski}}]{Udalski2015}
{Udalski}, A., {Szyma{\'n}ski}, M.~K., \& {Szyma{\'n}ski}, G. 2015, \actaa, 65, 1

\bibitem[{{Wambsganss}(1997)}]{1997MNRAS.284..172W}
{Wambsganss}, J. 1997, \mnras, 284, 172

\bibitem[{{Witt}(1990)}]{Witt1990}
{Witt}, H.~J. 1990, \aap, 236, 311

\bibitem[{{Witt} \& {Mao}(1994)}]{WittMao1994}
{Witt}, H.~J., \& {Mao}, S. 1994, \apj, 430, 505

\bibitem[{{Yee} \& {Gould}(2023)}]{Yee2023}
{Yee}, J.~C., \& {Gould}, A. 2023, arXiv e-prints, arXiv:2306.15037

\bibitem[{{Zhu} \& {Dong}(2021)}]{2021ARA&A..59..291Z}
{Zhu}, W., \& {Dong}, S. 2021, \araa, 59, 291

\end{thebibliography}





\end{document}